\documentclass[preprint,showpacs,preprintnumbers,amsmath,amssymb,aps,pre]{revtex4}

\usepackage[utf8]{inputenc} 
\usepackage{epsfig}
\usepackage{natbib}
\usepackage{amsmath}
\usepackage{times}
\usepackage{subfigure}
\usepackage{epstopdf} 
\usepackage{graphicx}
\usepackage{color}
\usepackage{pict2e}
\newcommand{\comment}[1]{}

\begin{document}

\title{Ergodic decay laws in Newtonian and relativistic chaotic scattering}

\author{Diego S. Fern\'{a}ndez}
\email{diego.sfernandez@urjc.es}
\author{\'{A}lvaro G. L\'{o}pez}
\author{Jes\'{u}s M. Seoane}
\author{Miguel A. F. Sanju\'{a}n}
\affiliation{Nonlinear Dynamics, Chaos and Complex Systems Group, Departamento de
F\'{i}sica, Universidad Rey Juan Carlos, Tulip\'{a}n s/n, 28933 M\'{o}stoles, Madrid, Spain}

\date{\today}

\begin{abstract}
In open Hamiltonian systems, the escape from a bounded region of phase space according to an exponential decay law is frequently associated with the existence of hyperbolic dynamics in such a region. Furthermore, exponential decay laws based on the ergodic hypothesis are used to describe escapes in these systems. However, we uncover that the presence of the set that governs the hyperbolic dynamics, commonly known as the chaotic saddle, invalidates the assumption of ergodicity. For the paradigmatic H\'{e}non-Heiles system, we use both theoretical and numerical arguments to show that the escaping dynamics is non-ergodic independently of the existence of KAM tori, since the chaotic saddle, in whose vicinity trajectories are more likely to spend a finite amount of time evolving before escaping forever, is not utterly spread over the energy shell. Taking this into consideration, we provide a clarifying discussion about ergodicity in open Hamiltonian systems and explore the limitations of ergodic decay laws when describing escapes in this kind of systems. Finally, we generalize our claims by deriving a new decay law in the relativistic regime for an inertial and a non-inertial reference frames under the assumption of ergodicity, and suggest another approach to the description of escape laws in open Hamiltonian systems.
\end{abstract}

\maketitle

\section{Introduction} \label{sec:1}

When a point mass travels freely and meets a region of interaction, commonly described in terms of a potential, its state of motion is irremediably affected. The particle may escape eventually from the influence of the potential, resuming its free journey until another obstacle is encountered. This physical phenomenon is known as a scattering process, and if the potential function is nonlinear, the particle might experience a chaotic scattering \cite{seoane2013}. In this case, the particle will be subject to chaotic dynamics that will cause its final state of motion to be extremely sensitive to modifications in the initial conditions, which hinders its predictability \cite{aguirre2009}. In most scattering processes, the particle experiences the effect of chaotic dynamics just for a finite amount of time before escaping, i.e., its motion undergoes a transient chaotic dynamics \cite{lai2010,tel2015}. This phenomenon is typical in open 2D time-independent Hamiltonian systems, among which the H\'{e}non-Heiles system constitutes a paradigmatic example \cite{aguirre2001}. Briefly, transient chaotic dynamics is governed by the presence in phase space of the chaotic saddle, also called invariant nonattracting chaotic set \cite{ott1993}, which rules how much time particles spend in the phase space region called the scattering region. Chaotic scattering is a fundamental field in nonlinear dynamics, which has many different applications in physics, such as molecule scattering, advection of particles in fluid mechanics, transition of materials, or even stars escaping from galaxies \cite{lin2013, daitche2014, toledomarin2018, navarro2019}.

To study the probabilistic laws that govern the particle escape from the influence of the potential, large ensembles of particles can be launched inside the scattering region, where they interact with the potential well. In this way, particles escape following an exponential decay law when hyperbolic chaos governs the dynamics, whereas escapes take place according to algebraic laws when the underlying dynamics is nonhyperbolic \cite{bauer1990, alt1996, lau1991, kokshenev2000, bunimovich2005}. On the other hand, exponential decay laws are sometimes associated with ergodic motions, so it is common to find works in which the concepts of chaos and ergodicity are used in an equivalent way \cite{zhao2007}. Boltzmann coined the terms \emph{ergoden} or \emph{ergodische}, whose Greek etymological origin means ``energy path" \cite{moore2015}, to refer to systems in which trajectories, if left to itself for long time enough, will pass close to nearly all the dynamical states on the energy surface \cite{lebowitz1973}, i.e., the dynamical states compatible with a constant energy. It is widely known that in small systems with a few degrees of freedom chaos is decisive to determine if the laws of statistical mechanics are satisfied, and therefore it is the key to many proofs that demonstrate that systems are ergodic \cite{berdichevsky1988}.

However, exponential decay laws based on the ergodic hypothesis, which we shall name \emph{ergodic decay laws} hereafter for simplicity, are inadequate to describe the rich phenomenology of open Hamiltonian systems, as confirmed by the numerical results presented in this work. The reason is that assuming this hypothesis in the open regime is an oversimplifying approximation for small exits, since it implies disregarding the existence of KAM tori or the chaotic saddle in the nonhyperbolic and hyperbolic dynamical regimes, respectively. As is well known, KAM tori prevent particles to escape exponentially for long periods of time due to the phenomenon of stickiness, by which particles starting in a chaotic region can stick to the vicinity of the boundary of a regular region before escaping \cite{hillermeier1992, lai1992, mackay1992, contopoulos2010, bunimovich2012}. Therefore, the KAM stickiness clearly disallows the assumption of equiprobability of the phase space states, avoiding ergodic motions \cite{lebowitz1973, zheng1995, berdichevsky1991}.

On the other hand, whereas the presence of the chaotic saddle can make the decay law exponential, it is also responsible for the existence of particles that evolve and spend a finite amount of time near its vicinity before escaping. Specifically, a significant number of particles approach the chaotic saddle following its stable manifold, and move away from its vicinity following its unstable manifold \cite{ott1993}. In this manner, despite the fact that hyperbolic chaos rules the dynamics in these situations, the particle's motion is non-ergodic before escaping because the saddle vicinity occupies only a subset of the energy shell. Then, once again, we can not expect equiprobability of the phase space states, when following any escaping trajectory compatible with some value of the energy. Briefly speaking, in terms of measure theory, the measure that characterizes the escaping process from the chaotic saddle along the so-called unstable manifold is very different from the natural measure of a hyperbolic and closed Hamiltonian system \cite{altmann2008} that is ergodic and, therefore, that can be described by the microcanonical ensemble throughout the chaotic sea in its phase space \cite{zheng1995}.

One of the purposes of the present work is to extend ergodic decay laws to relativistic chaotic scattering, to study numerically their scope of applicability regarding the system's dynamical regimes, and to propose other ways to widen these laws taking into account statistical methods that do not rely on the ergodic hypothesis. Firstly, we provide a complete description of the H\'{e}non-Heiles model in Sec.~\ref{sec:2}, including striking aspects of escapes, such as the critical time. In Sec.~\ref{sec:3} we thoroughly discuss the concept of ergodicity in open Hamiltonian systems, and provide numerical evidence of the limited applicability of ergodic decay laws at the end of the section. Furthermore, we derive an analytic ergodic decay law in a relativistic version of the H\'{e}non-Heiles system in Sec.~\ref{sec:4}. We also derive this analytical decay law considering that escapes are measured by a noninertial reference frame comoving with a particle that never escapes from the scattering region. Finally, in Sec.~\ref{sec:5}, we conclude this work with a discussion of our findings and suggesting possible ways to obtain a more accurate characterization of decay laws utilizing statistical mechanics.

\section{Model description} \label{sec:2}

\begin{figure*}[htp!]
	\centering
	
	\includegraphics[width=0.31\textwidth]{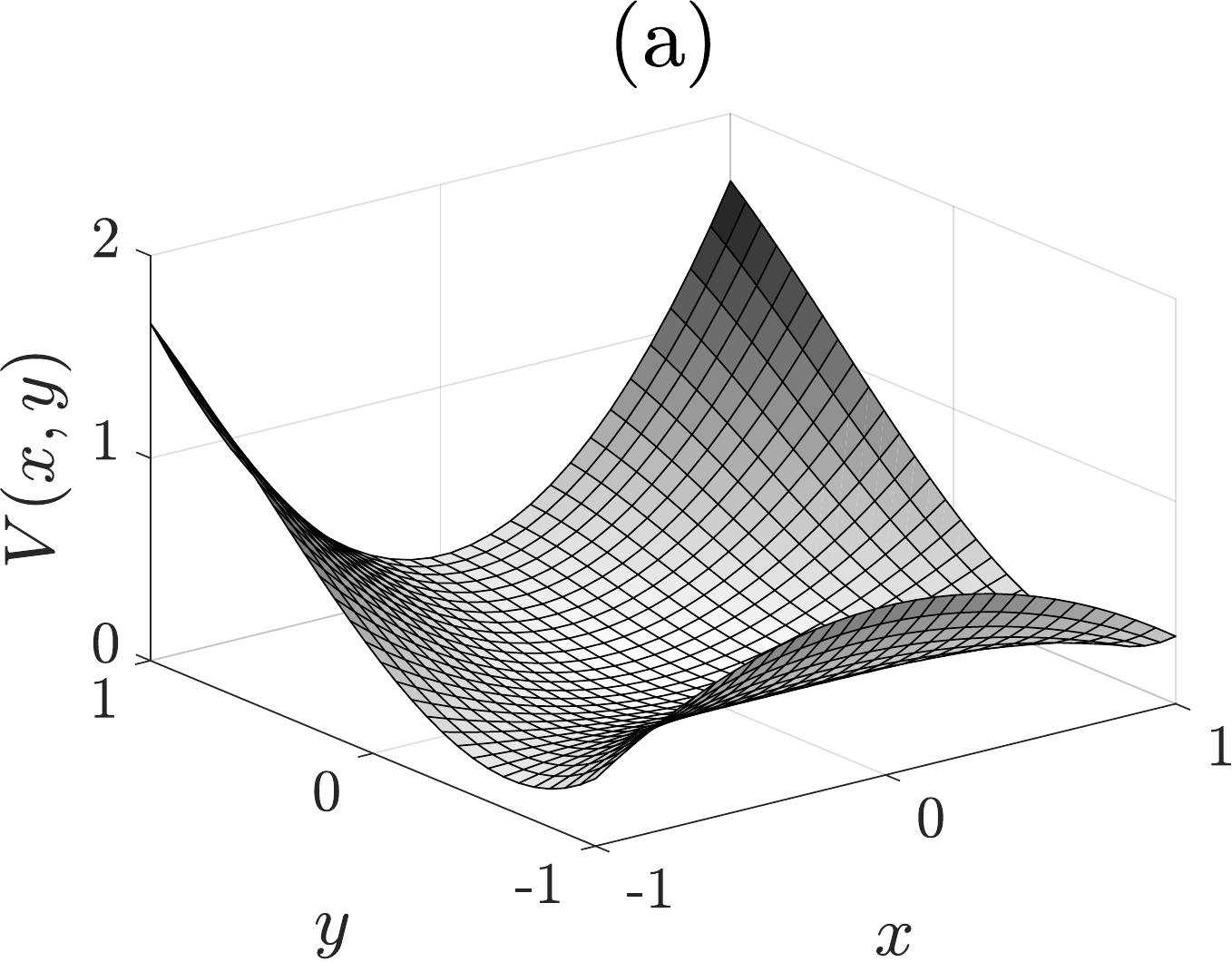}
	\quad	
	\includegraphics[width=0.31\textwidth]{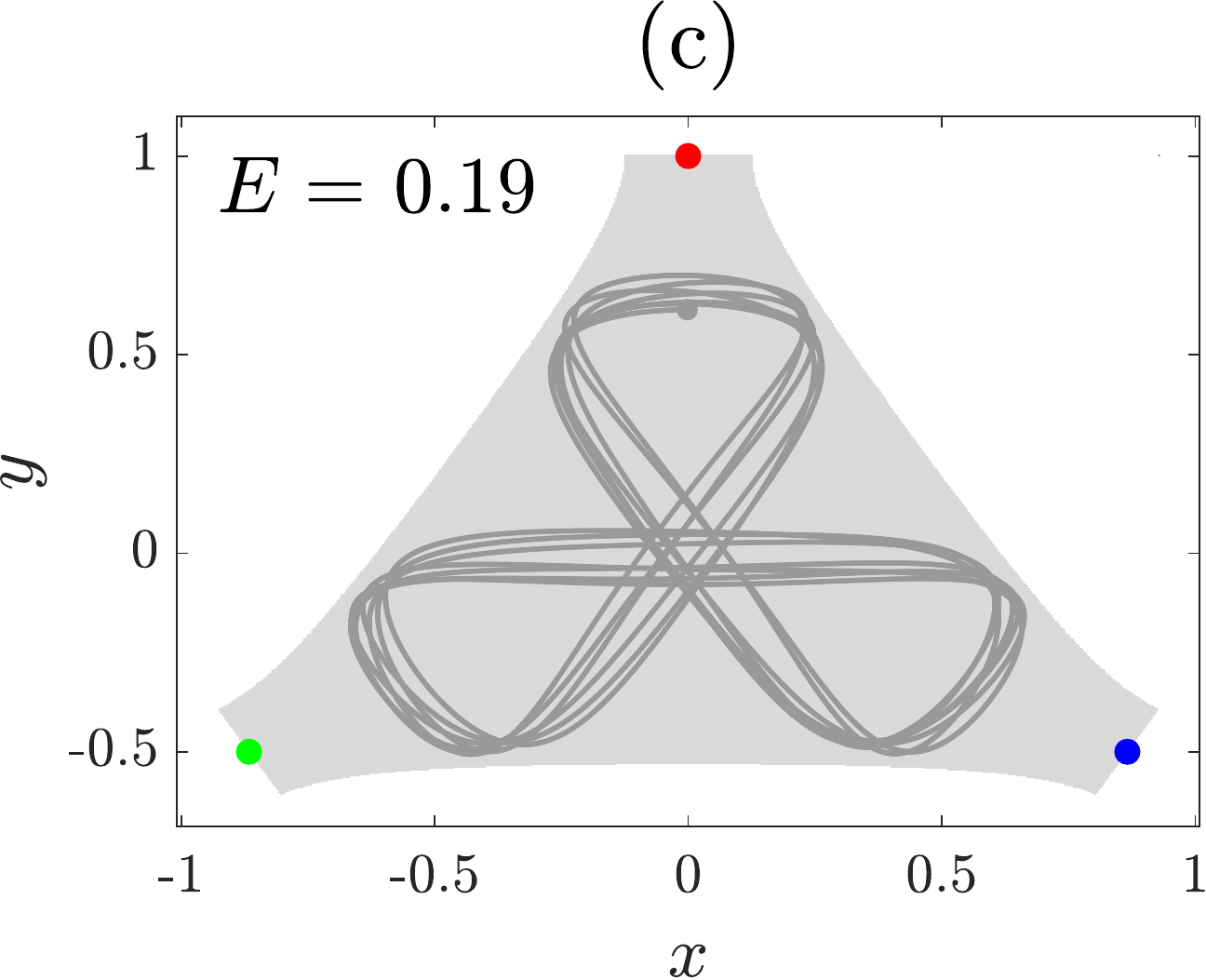}
	\quad
	\includegraphics[width=0.31\textwidth]{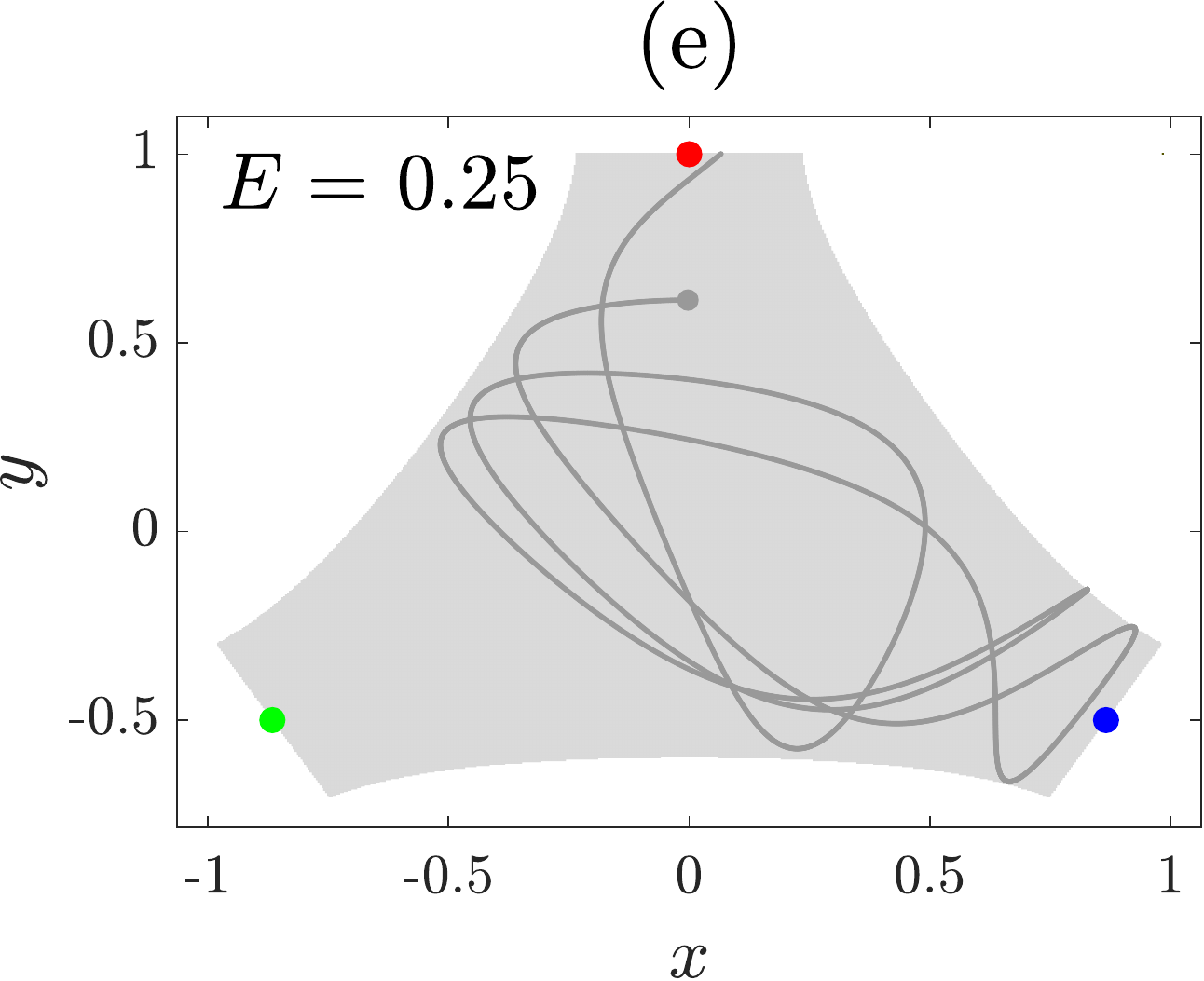}\\
	\bigskip
	\includegraphics[width=0.31\textwidth]{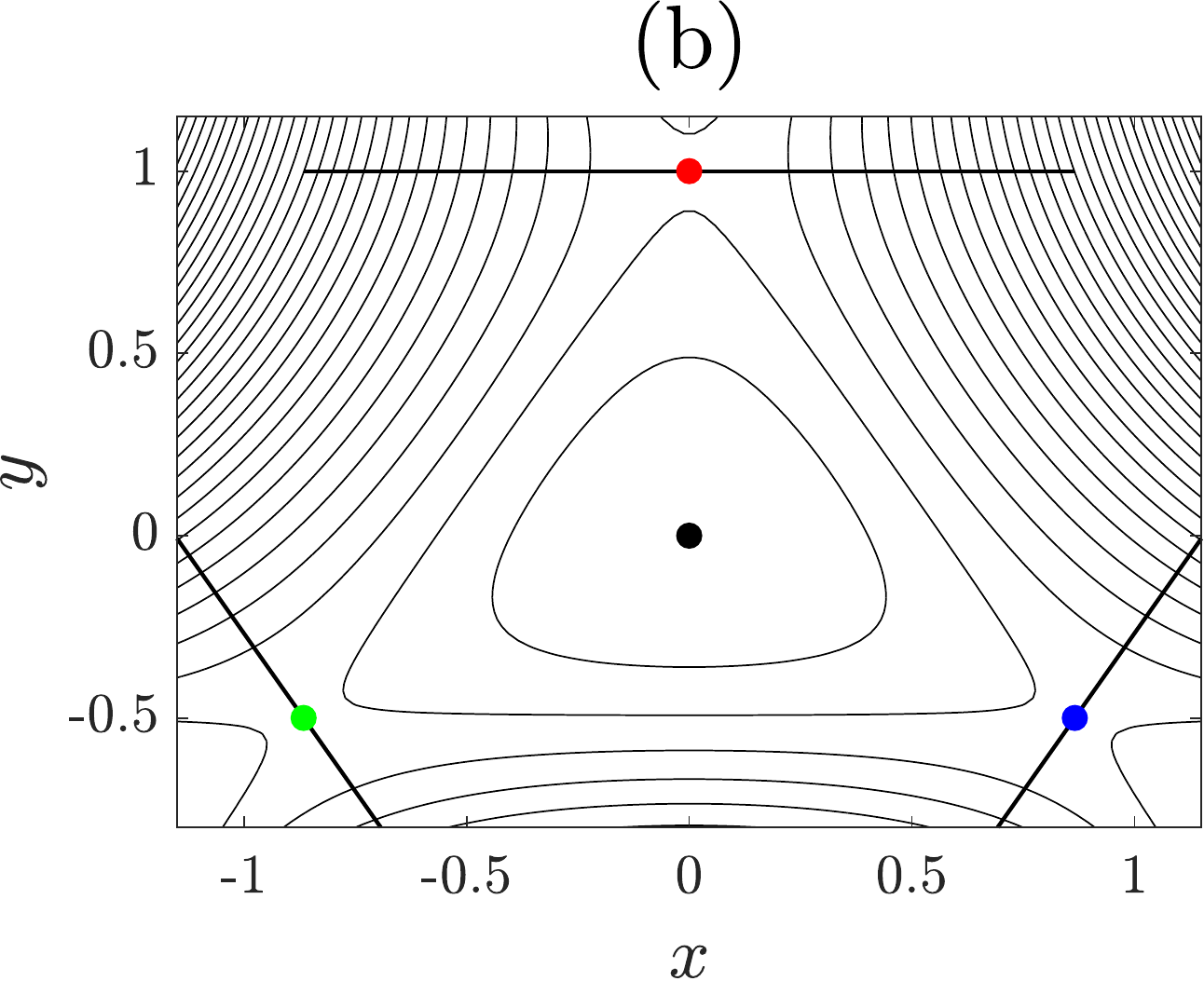}
	\quad
	\includegraphics[width=0.31\textwidth]{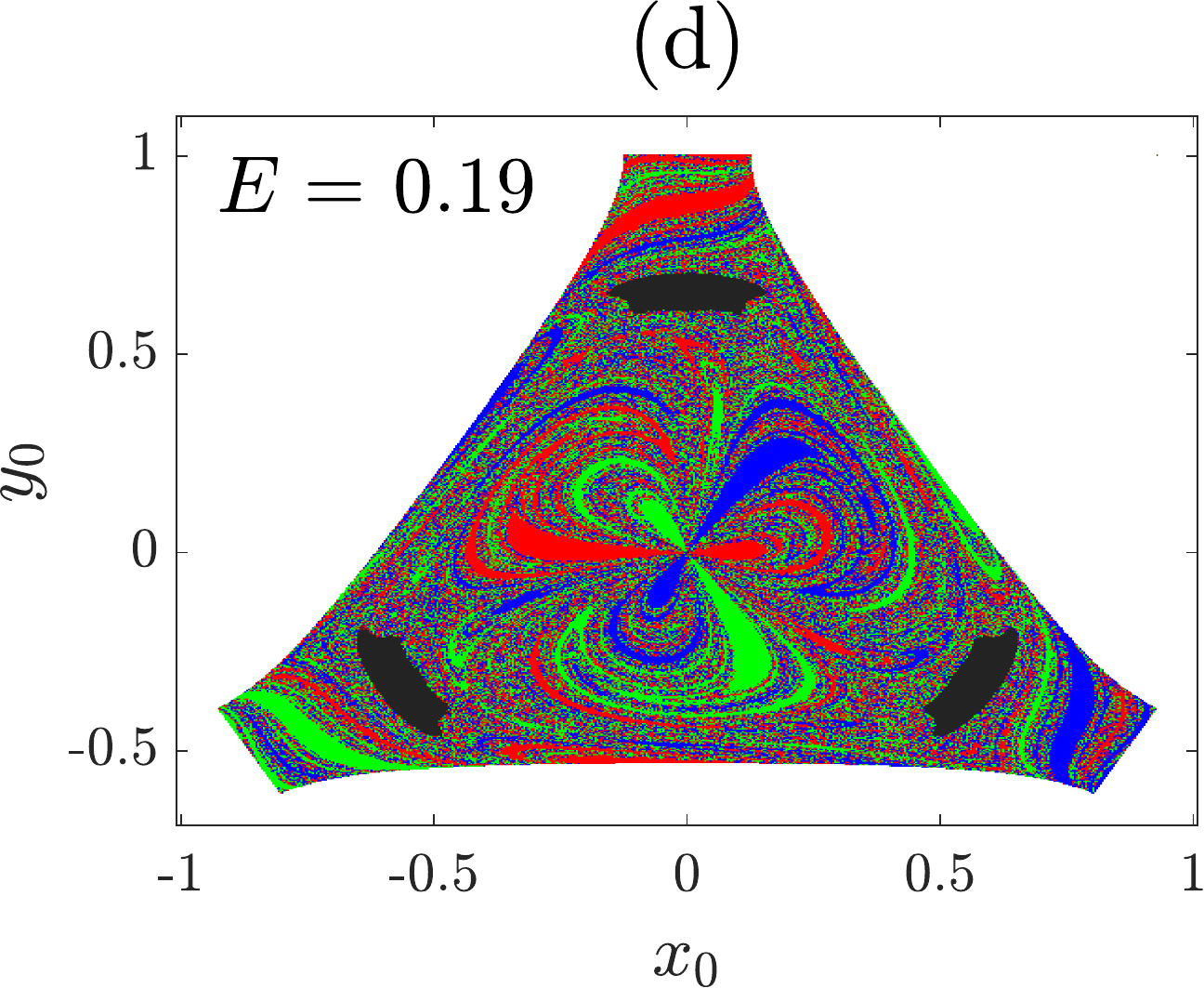}
	\quad
	\includegraphics[width=0.31\textwidth]{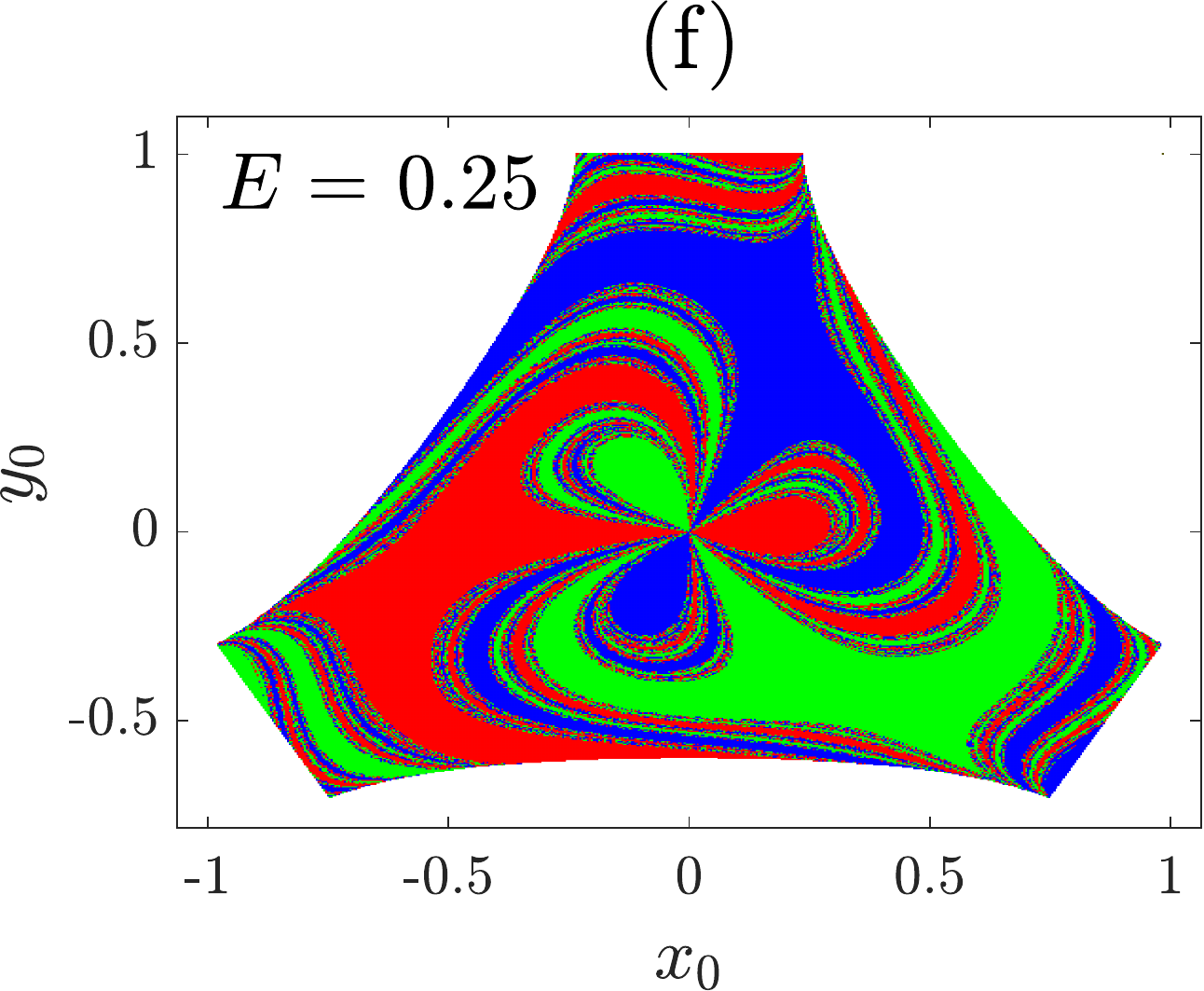}
	
	\caption{(a) H\'{e}non-Heiles potential, $V(x,y) = \frac{1}{2} (x^2 + y^2) + x^2y - \frac{1}{3} y^3$. (b) Closed and open isopotential curves projected into the configuration space $(x,y)$. We arbitrarily establish the exits of the potential well at the saddle points: Exit 1 (red), Exit 2 (green) and Exit 3 (blue). In this way, the scattering region is the allowed region of the potential well delimited by the saddle points lines. The potential minimum is located right at the center of the well (black dot). (c) Scattering region for $E = 0.19$ (light gray) and a particle trapped in a KAM structure describing a non-escaping orbit typical of the nonhyperbolic regime. (d) Exit basins and non-escaping orbits in black associated with $E = 0.19$, as computed by means of the tangential shooting method \cite{aguirre2001} with a 500$\times$500 resolution. (e) Similarly, for $E = 0.25$, an escaping orbit starting from the same initial condition that in (c), but in the hyperbolic regime. (f) Exit basins for $E = 0.25$.}
	\label{fig:1}
\end{figure*}

The H\'{e}non-Heiles system was introduced in 1964 to study the existence of a third integral of motion in models of galaxies with axial symmetry \cite{henon1964}. We  utilize a dimensionless and conservative version of the H\'{e}non-Heiles system where the particle dynamics is governed by the Hamiltonian \begin{equation} H(p_x,p_y,x,y) = \frac{p_x^2 + p_y^2}{2} + V(x,y), \label{eq:1} \end{equation} where $p_x$ and $p_y$ are the momenta, and $V(x,y)$ is a nonlinear potential function which depends on $x$ and $y$. In the H\'{e}non-Heiles system, the potential well is written as \begin{equation} V(x,y) = \frac{1}{2}(x^2 + y^2) + x^2 y - \frac{1}{3}y^3. \label{eq:2} \end{equation} The value of the mechanical energy is determined by the Hamiltonian, which is a conserved quantity, $H(p_x,p_y,x,y) = E$. When the energy is above a threshold value called the escape energy, $E_e = 1/6$, most trajectories are unbounded so that the system enters an open regime, where escapes are allowed and the potential well exhibits three symmetric exits, as shown in Fig.~\ref{fig:1}. In addition, we define the quantity $\Delta E \equiv E - E_e$ for convenience in computing decay laws in next sections.

The corresponding equations of motion are given by \begin{equation} \begin{aligned} \dot{x} = & \frac{\partial H}{\partial p_x} = p_x, & \dot{p_x} = -\frac{\partial H}{\partial x} =-x-2xy, \\ \dot{y} = & \frac{\partial H}{\partial p_y} = p_y, & \dot{p_y} = -\frac{\partial H}{\partial y} = y^2 -x^2 -y. \end{aligned} \label{eq:3} \end{equation} The fixed points of the system can be computed from these latter equations. In this manner, we see that the system has three fixed points located at the saddle points of the potential well, namely, $(x_s, y_s) = (0,1)$, $(\sqrt{3}/2,-1/2)$ and $(-\sqrt{3}/2,-1/2)$, and another fixed point right in the minimum of the well, $(x_m, y_m) = (0,0)$. In addition, there exist three highly unstable periodic orbits known as Lyapunov orbits \cite{contopoulos1990}. If a trajectory crosses through one of them, it escapes towards infinity and never returns back to the scattering region. The Lyapunov orbits are placed extremely close to the saddle points in the present version of the H\'{e}non-Heiles system, and therefore it is convenient to define the scattering region as the allowed region of the potential well delimited by the lines that cross the saddle points, as can be visualized in Fig.~\ref{fig:1}(b). Unless otherwise specified, trajectories have been computed by means of an adaptive Runge-Kutta-Fehlberg method \cite{burden2015} with a relative tolerance of $10^{-12}$ for each solution of the system of equations and a maximum value of the integration step of $5 \cdot 10^{-3}$, small enough to conserve the mechanical energy along the trajectory.

The underlying dynamics of open systems is determined by the presence of KAM tori in phase space. They are sticky sets of quasiperiodic orbits that may delay escapes or even trap particles forever within the scattering region, as shown in Fig.~\ref{fig:1}(c). Then, there exist two open dynamical regimes. On the one hand, the nonhyperbolic regime where KAM tori coexist with the chaotic saddle and the phase space exhibits regions where the dynamics is regular surrounded by a chaotic sea \cite{sideris2006}. An example of the typical nonhyperbolic topology of the exit basins is presented in Fig.~\ref{fig:1}(d). As already indicated, the KAM tori stickiness may significantly influence the global properties of the dynamics, such as the decay law, which can become algebraic at long times.
	
On the other hand, if the system energy is increased, the KAM tori that dominate the dynamics disappear and thus the hyperbolic regime begins. The destruction of such KAM structures has been studied recently, yielding approximately an energy value of $E \approx 0.2309$ for the transition between dynamical regimes \cite{nieto2020}. After this energy value, the chaotic saddle dominates the dynamics, making the system chaotic (see, e.g., Figs.~\ref{fig:1}(e) and \ref{fig:1}(f)). Nonetheless, some remnant KAM tori may survive for $E>0.2309$, after the destruction of the last relevant KAM tori. The presence of such remnant structures in the hyperbolic regime show that purely hyperbolic dynamics is unlikely to exist \cite{duarte1994}. However, they have been extensively studied and found to be extremely small and difficult to locate \cite{barrio2009, barrio2020}, so that their possible stickiness barely affects the decay law: an ensemble of particles escapes conforming an exponential law at long times in the hyperbolic regime. Finally, this situation is different in area-preserving maps, where there exist apparently relevant sticky KAM structures for ranges of parameter values arbitrarily large \cite{miguel2015}.

\begin{figure*}[htp!]
	\centering
	
	\includegraphics[width=0.485\textwidth]{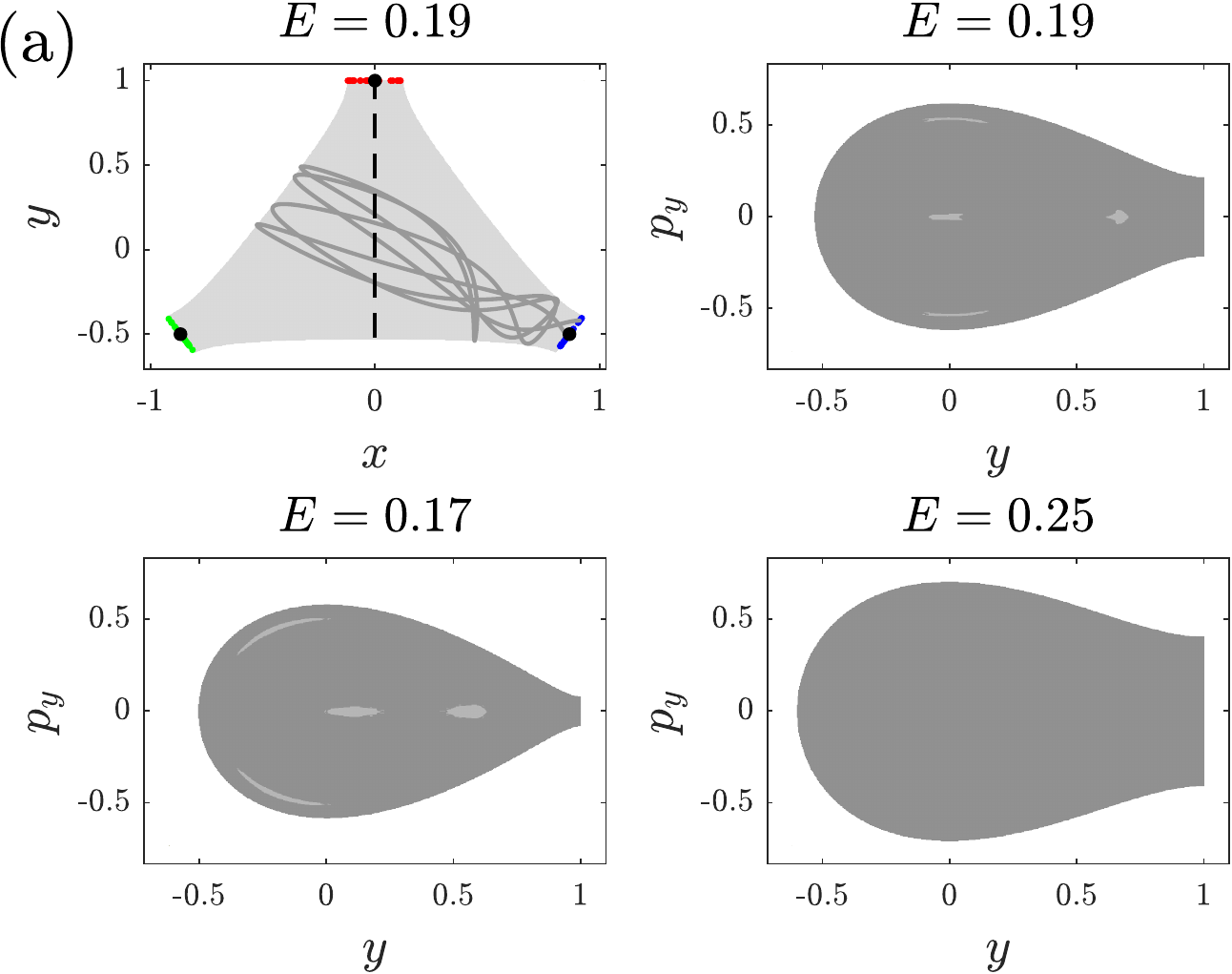}
	\quad
	\includegraphics[width=0.4855\textwidth]{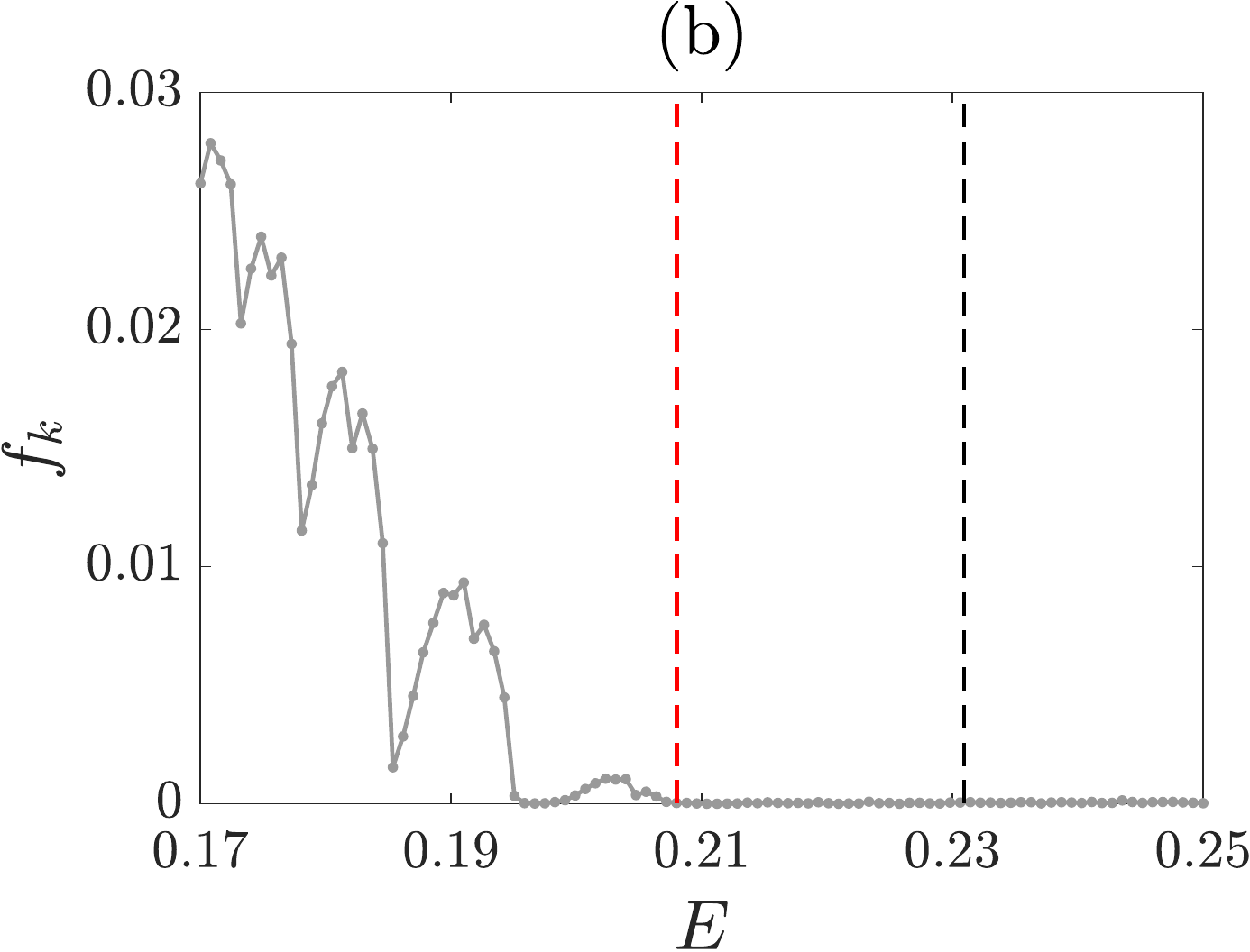}
	
	\caption{Computing the fraction of KAM tori in the surface of section $(y,p_y)$ placed at $x = 0$, where only escaping trajectories are launched from the exits of the scattering region (more details in the text and in Appendix). (a) For $E = 0.19$, a trajectory launched from the Exit 3 with eight crossings though the surface of section before escaping through Exit 3, again. We also show three examples of surfaces of section for $E = 0.17$, $0.19$ and $0.25$, where the mapped regions (darker gray) and KAM tori regions (lighter gray) are depicted. (b) Fraction of the surface of section occupied by KAM tori $f_k$ versus the system energy, specifically, 50 equally spaced values belonging to the interval $E \in [0.17,0.25]$. The dashed lines are placed at $E = 0.208$ (red) and $0.2309$ (black), respectively. We have set a 500$\times$500 grid resolution for the surface of section and $10^6$ initial conditions have been launched for each energy value.}
	\label{fig:2}
\end{figure*}

In order to illustrate how the volume in phase space occupied by the KAM tori evolves as the system energy is varied, we analyze the well-known surface of section $(y,p_y)$ located at $x = 0$, which is a convenient plane transverse to the orbits. Similar surfaces of section have been extensively used in area-preserving maps \cite{lai1992} and in continuous-time systems as well \cite{aguirre2001}. Thus, particles are launched from the exits of the scattering region with random initial shooting angles into its interior. These trajectories map the entire surface of section except the area occupied by the KAM tori. Using a partition of the surface of section with high resolution, it is easy to compute what fraction of points have not been visited by the escaping trajectories, which is the fraction occupied by the KAM tori (see Appendix at the end of the manuscript for some method specifications). We show in Fig.~\ref{fig:2}(a) examples of which regions are mapped (darker gray) and those occupied by the KAM tori (lighter gray) for three values of the energy, and in Fig.~\ref{fig:2}(b) the fraction of KAM tori versus the system energy. The shape of the curve is non-trivial due to the complexity of the KAM tori destruction \cite{barrio2020}. Although the volume of the phase space occupied by tori KAM is residual for $E > 0.208$ (dashed red line), it has been shown that KAM structures can influence up to $E \approx 0.2309$ still some global properties of the dynamics, such as the unpredictability of the final state of the system \cite{nieto2020}. For this reason, along this work, we set the hyperbolic regime in the energy range $E > 0.2309$.

In addition to KAM tori, a relevant set of the phase space is the chaotic saddle. Specifically, it is an invariant set with zero-Lebesgue measure resulting from the intersection of the stable and unstable manifolds. The stable (unstable) manifold consists of a self-similar fractal set of an uncountable number of orbits that approach the saddle as $t \to \infty$ ($t \to -\infty$) \cite{ott1993_2}. On the other hand, the stable manifold can be understood as the boundary of the exit basins, which exhibits the Wada property in the H\'{e}non-Heiles system \cite{aguirre2001}. For this reason, the stable manifold of the chaotic saddle (and therefore itself) can be characterized by means of the fractal dimension of the exit basins boundary. As the energy is increased, such boundary becomes smoother, the fractal dimension of the chaotic saddle decreases and the system becomes more predictable \cite{nieto2020}.

\begin{figure*}[b!]
	\centering
	
	\includegraphics[width=0.485\textwidth]{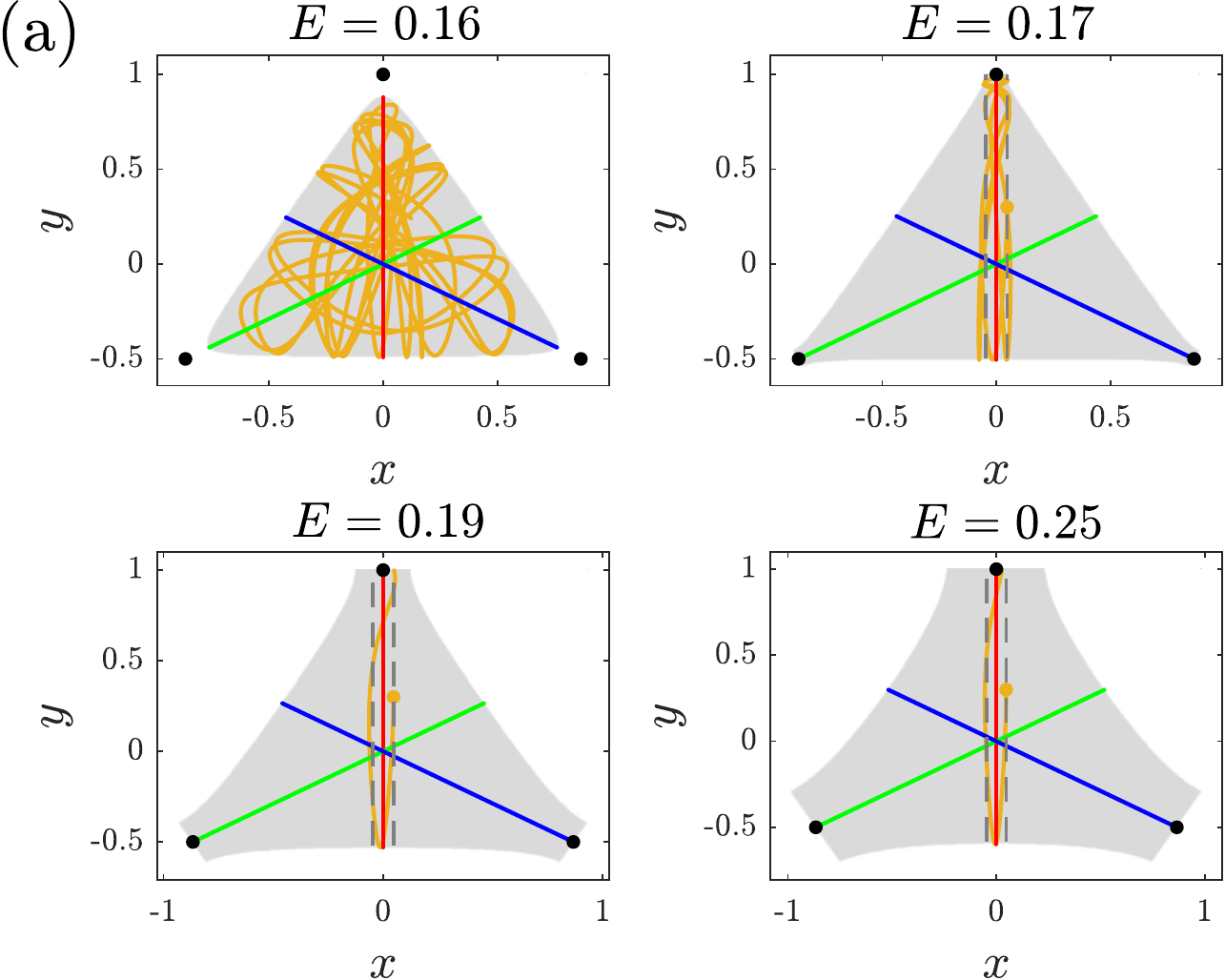}
	\quad
	\includegraphics[width=0.485\textwidth]{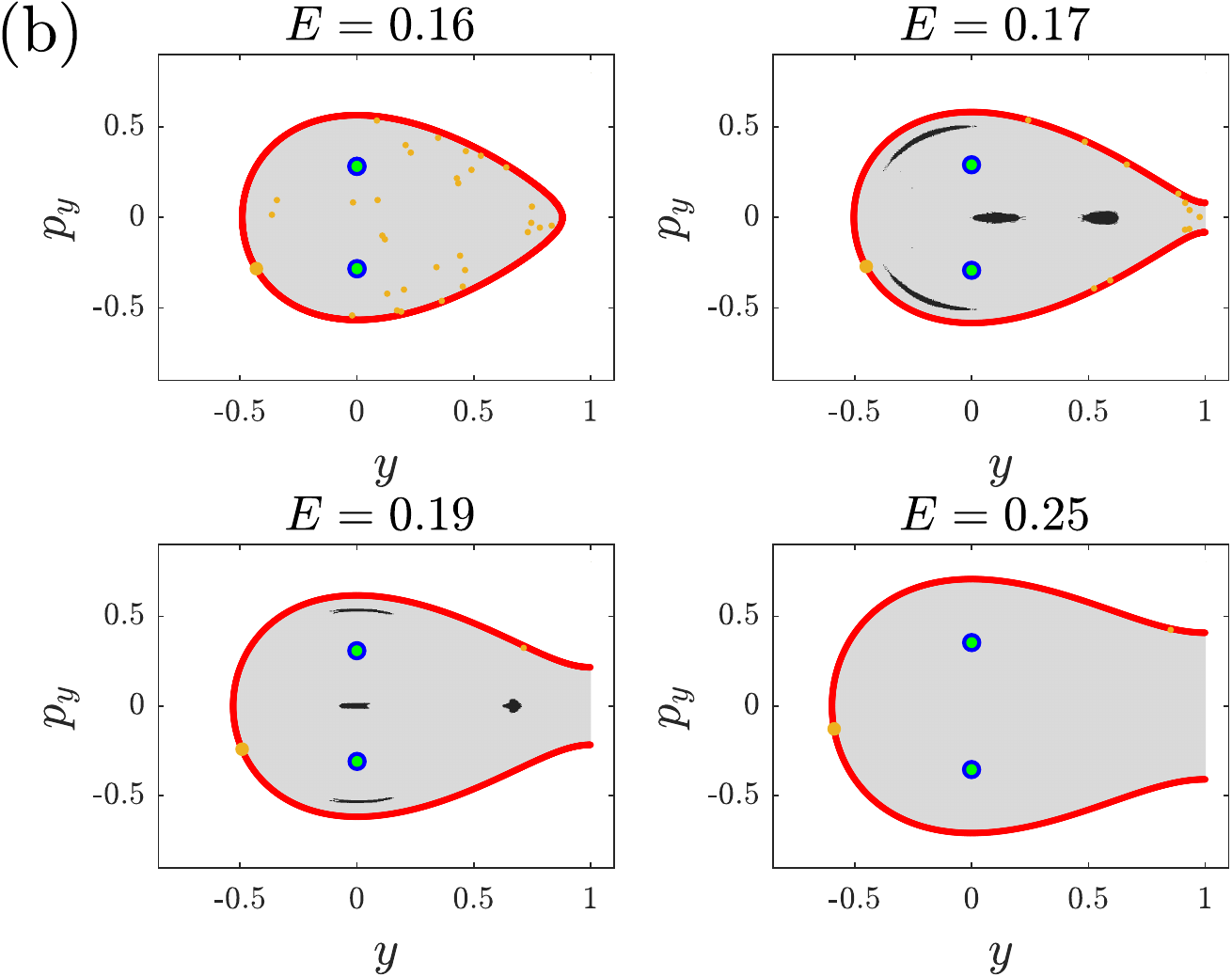}\\
	\bigskip
	\includegraphics[width=0.485\textwidth]{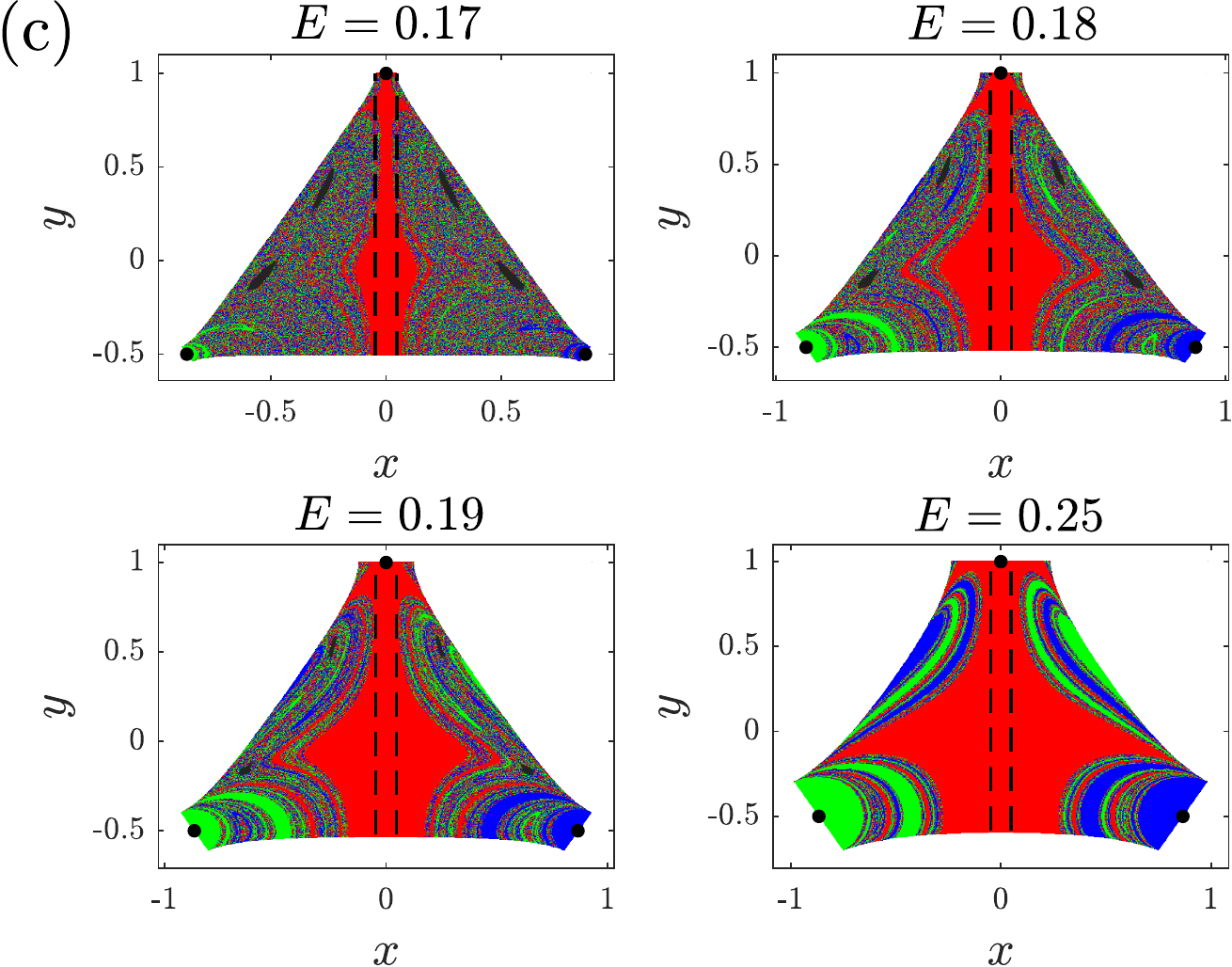}
	\quad
	\includegraphics[width=0.475\textwidth]{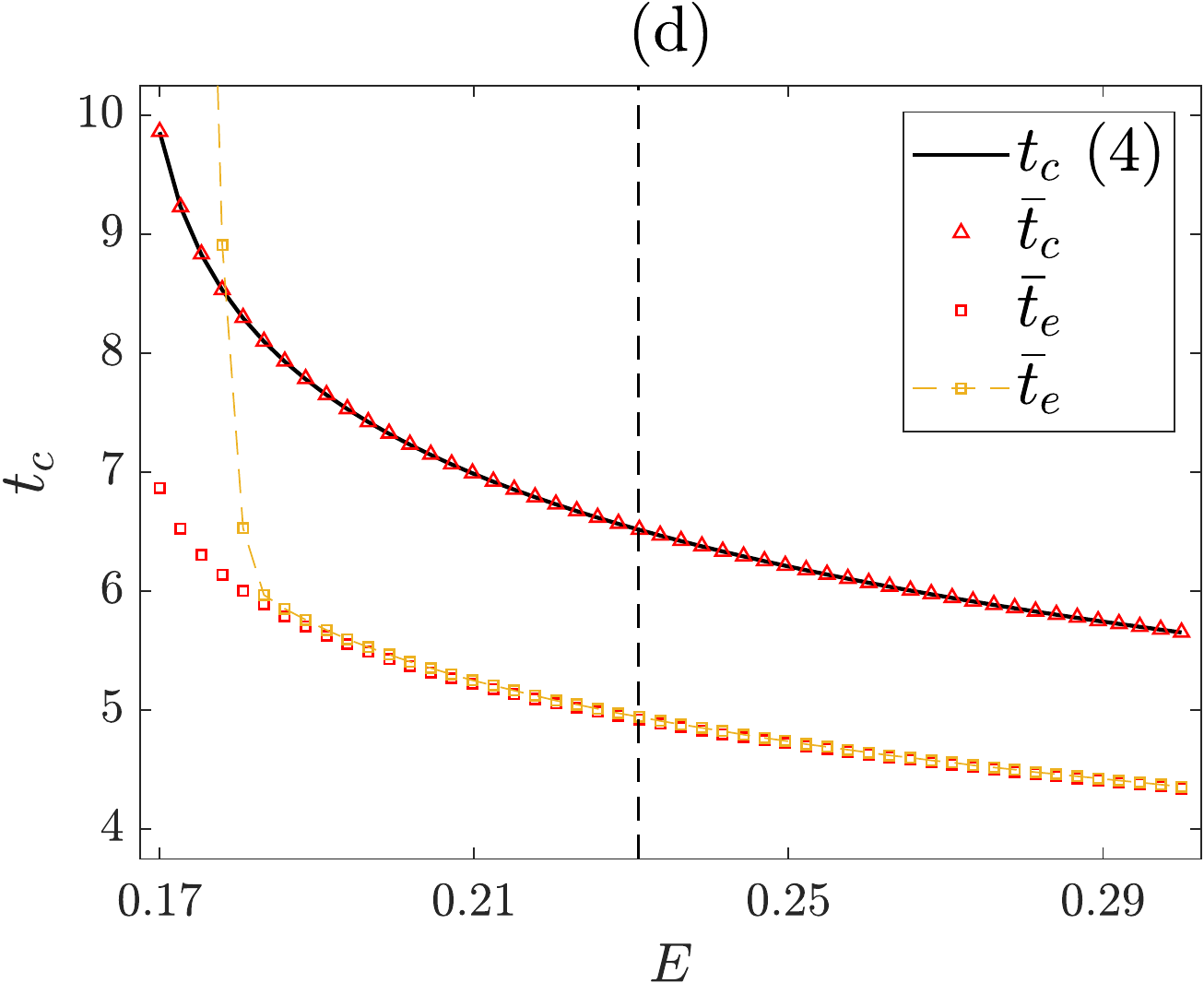}
	
	\caption{(a) Critical trajectories (red, green and blue) in the configuration space for $E = 0.16$ (closed regime) and $E = 0.17$, $0.19$ and $0.25$ (open regime). We also simulate another trajectory (yellow) initialized near the critical trajectory associated with the Exit 1. (b) Critical trajectories in the space $(y,p_y)$ when $x = 0$. They allow to visualize the presence of KAM tori (black) and their proximity to the critical trajectories. (c) Exit basins as computed by shooting particles always with a constant initial angle, $\theta_0 = 3\pi /2$. (d) Critical times versus the system energy as a numerical result of Eq.~\eqref{eq:4} (black line) and as an average from the escape times of three critical trajectories (red triangles). On the other hand, $\bar{t}_e$ (red squares) represents the average escape times of $5 \cdot 10^5$ particles launched randomly along the critical trajectory associated with the Exit 1, and $\bar{t}_e$ (yellow) means the average escape times of particles launched with $\theta_0 = 3\pi /2$ and randomly distributed within the delimited region by the dashed lines depicted in (a) and (c). Finally, the vertical dashed line splits the nonhyperbolic and hyperbolic regimes at an approximate energy value $E = 0.2309$.}
	\label{fig:3}
\end{figure*}

An important concept, such as the critical time, has not been addressed in this system yet. The critical time is defined as the escape time spent by a particle launched from a saddle point of the potential towards the center of the well \cite{bolotin2008, bolotin2010}. More specifically, it is the maximum time that straight-line trajectories can spend inside the scattering region before escaping. We name these trajectories as the critical trajectories and there exist as many as saddle points the potential has (see Fig.~\ref{fig:3}(a)). We compute this time in the H\'{e}non-Heiles system utilizing the critical trajectory associated with Exit 1, \begin{equation} t_c(E) = 2 \int_{y_s}^{y_f} \frac{dy}{p} = 2 \int_{1}^{y_f} \frac{dy}{\sqrt{2E - y^2 + \frac{2}{3}y^3}}, \label{eq:4}\end{equation} where $y_s = 1$ is the $y$-coordinate of the saddle point located at the Exit 1, and $y_f$ is the $y$-coordinate of the point of the configuration space which satisfies the equality $E = V(0,y_f)$. Finally, $p = \sqrt{p_x^2 + p_y^2}$ is the modulus of the particle momentum.

Interestingly, critical trajectories are not disturbed by the presence of KAM tori in light of Fig.~\ref{fig:3}(b). In addition, already in Fig.~\ref{fig:3}(a), we also show the critical trajectories even when the potential well is closed, for $E = 0.16$. In this example, the three trajectories remain bounded along straight lines over 100 time units, bouncing back and forth against the potential barriers. They can be simulated over any arbitrarily longer time and would describe the same straight-line bounded behavior, because the so-called critical trajectories are closely related to some of the stable normal modes of the H\'{e}non-Heiles system in its closed regime \cite{barrio2009,barrio2020}. More examples of critical trajectories are shown when the system is open in Fig.~\ref{fig:3}(a) again. Therefore, as the critical trajectories are not deviated by the likely influence of KAM tori or the chaotic saddle, we can safely state that critical trajectories are the orbits in the open regime that spend the longest time without suffering any effect of the sets that govern the dynamics. In order to demonstrate the latter, we compute the exit basins shown in Fig.~\ref{fig:3}(c) by shooting particles with a constant initial angle $\theta_0 = 3\pi /2$. We notice that there exists a smooth region of initial conditions (depicted in red) near the Exit 1 critical trajectory, that is surrounded by the typical fractal regions produced by the influence of the sets that governs the dynamics, where the final state of particles is subject to chaos. As this smooth region becomes larger as the system energy increases, it implies that a significant amount of particles starting from it might escape having only suffered negligible effects from such mentioned sets on their trajectories and their escape times.

The critical times versus the system energy as a numerical result of Eq.~\eqref{eq:4} and as an average computed from the escape times of three critical trajectories are in perfect agreement as shown in Fig.~\ref{fig:3}(d). As pointed out above, the values of critical times are not disturbed by the presence of KAM tori. Furthermore, we compute the average escape times of two different ensembles of particles. On the one hand, we randomly launch particles along the Exit 1 critical trajectory, i.e., with $x_0 = 0$ and $\theta_0 = 3 \pi / 2$ as initial conditions. As expected, the average escape times computed are always less that the critical times, since simulated trajectories are just shorter parts of the Exit 1 critical trajectory. On the other hand, we simulate particles with $\theta_0 = 3\pi /2$ again, but randomly distributed within the region delimited by the limits of the Exit 1 for $E = 0.17$, i.e., $x_0 = \mp \sqrt{2 \Delta E /3}$, where $\Delta E = 0.17 - 1/6$ (see the dashed black lines in Fig.~\ref{fig:3}(a) or \ref{fig:3}(c)). Simulations reveals that average escape times are less than critical times for approximately $E > 0.18$ because the majority of these particles bounce against the potential barrier and escape through the Exit 1, having evolved shorter trajectories that critical trajectories in general. We recall that all these particles belong to the smooth region mentioned above.

Finally, according to the results about critical times, we state that all escaping orbits after the critical time are affected by KAM tori and the saddle in the nonhyperbolic and hyperbolic regimes, respectively. Conversely, as indicated above, there exist some escaping orbits that are negligibly affected by these sets before the critical time. Importantly, this fact is reflected in the decay curves: a sudden change of the tendency of the escapes occurs when the critical time is achieved. For instance, escapes take place more slowly after the critical time when the dynamics is nonhyperbolic because all escaping orbits are affected by KAM tori stickiness, as shown in the next section.

\section{Decay law in the Newtonian H\'{e}non-Heiles system} \label{sec:3}

Theoretically, a Hamiltonian system is ergodic when its energy surface uniquely consists of a chaotic sea, due to the existence of a dense set of highly unstable periodic orbits densely embedded in it \cite{ott1993_2}. However, a problem may arise when testing if the motion is ergodic or not in open systems, since it is necessary to simulate trajectories trapped in the scattering region for a large amount of time. For example, in open systems, simulated trajectories generally escape before accomplishing long enough times to fill densely the energy surface and test the assumption of ergodicity. In this regard, a modified H\'{e}non-Heiles system with reflecting walls, where elastic collisions occur, was proposed to prevent particles from escaping \cite{zheng1995}. The authors provide numerical results claiming that this modified H\'{e}non-Heiles system behaves ergodically in the hyperbolic regime when the walls are placed far from the origin of the potential well. Thus, one might think that if the reflecting walls are placed infinitely far from the origin, the H\'{e}non-Heiles system would be recovered and, following the computational results shown above, one would obtain that the open and hyperbolic H\'{e}non-Heiles system behaves ergodically. However, this limit case is unattainable in computational practice, since the H\'{e}non-Heiles system is open and does not allow any escaped trajectory to return to the scattering region again. As we will discuss below, the fact that trajectories escape forever causes the escaping process to be non-ergodic.

Since the chaotic saddle is a zero-Lebesgue-measure set and the definition of ergodicity only requires that almost all trajectories in the energy surface fill it densely \cite{moore2015}, the presence of such an invariant set should not hinder a priori the application of the ergodic hypothesis. However, saddle sets separate the phase space into disjoint sets, what invalidates the ergodic hypothesis. This is evinced by the fact that a trajectory launched directly towards any of the exits can leave the scattering region without returning (for instance, see any critical trajectory). Moreover, the phase space fragmentation is manifested by the existence of large and differentiated exit basins, as can be appreciated in Fig.~\ref{fig:1}(f). Therefore, the approximation of ergodicity to derive decay laws can only be made for energies close to the energy escape, i.e., when the exits are vanishingly small and the exit basins exhibit a completely fractalized structure. Furthermore, and as we show numerically right ahead, this hypothesis also demands to consider short enough times. Finally, as is well known, we recall that the stickiness and the possible entrapment of the particle developed by KAM tori prevent that a set of trajectories with a Lebesgue measure different from zero satisfies the ergodic hypothesis \cite{lebowitz1973, zheng1995, berdichevsky1991}.

Having stated the problems of assuming the ergodic hypothesis in open systems, we elucidate below the range of applicability of an ergodic decay law to describe exponential escapes. At first, the evolution of the number of particles inside the scattering region for any hyperbolic system can be expressed without loss of generality as \begin{equation} N(t) = N_0 e^{-\alpha(E) t}, \label{eq:5} \end{equation} where $\alpha(E)$ is the decay rate and depends on the mechanical energy of the system. A decay law, $\alpha_e$, has been derived analytically in the case of the open H\'{e}non-Heiles system \cite{zhao2007}, obtaining \begin{equation} \alpha_e (\Delta E) = \frac{\sqrt{3} \Delta E}{S(\Delta E)}, \label{eq:6}\end{equation} where the function $S(\Delta E)$ is the area of the scattering region as defined in Sec.~\ref{sec:2}, which can be computed by utilizing a Monte Carlo method. The calculus of the ergodic decay rate, $\alpha_e$, is accomplished by means of the microcanonical ensemble, which is the only ensemble able to describe the statistical properties of ergodic systems \cite{gutzwiller1990}.

\begin{figure*}[b!]
	\centering
	
	\includegraphics[width=0.463\textwidth]{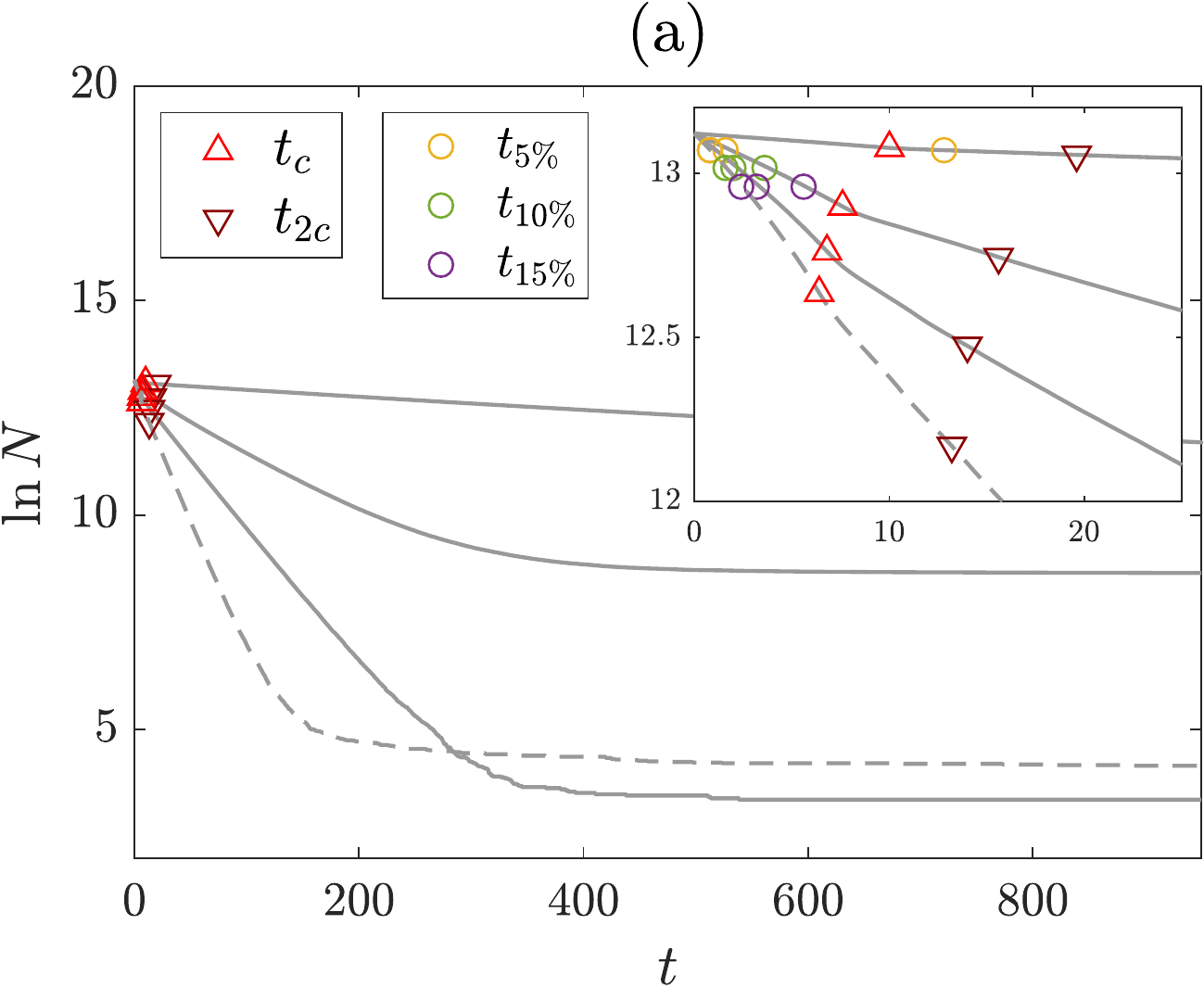}
	\quad
	\includegraphics[width=0.475\textwidth]{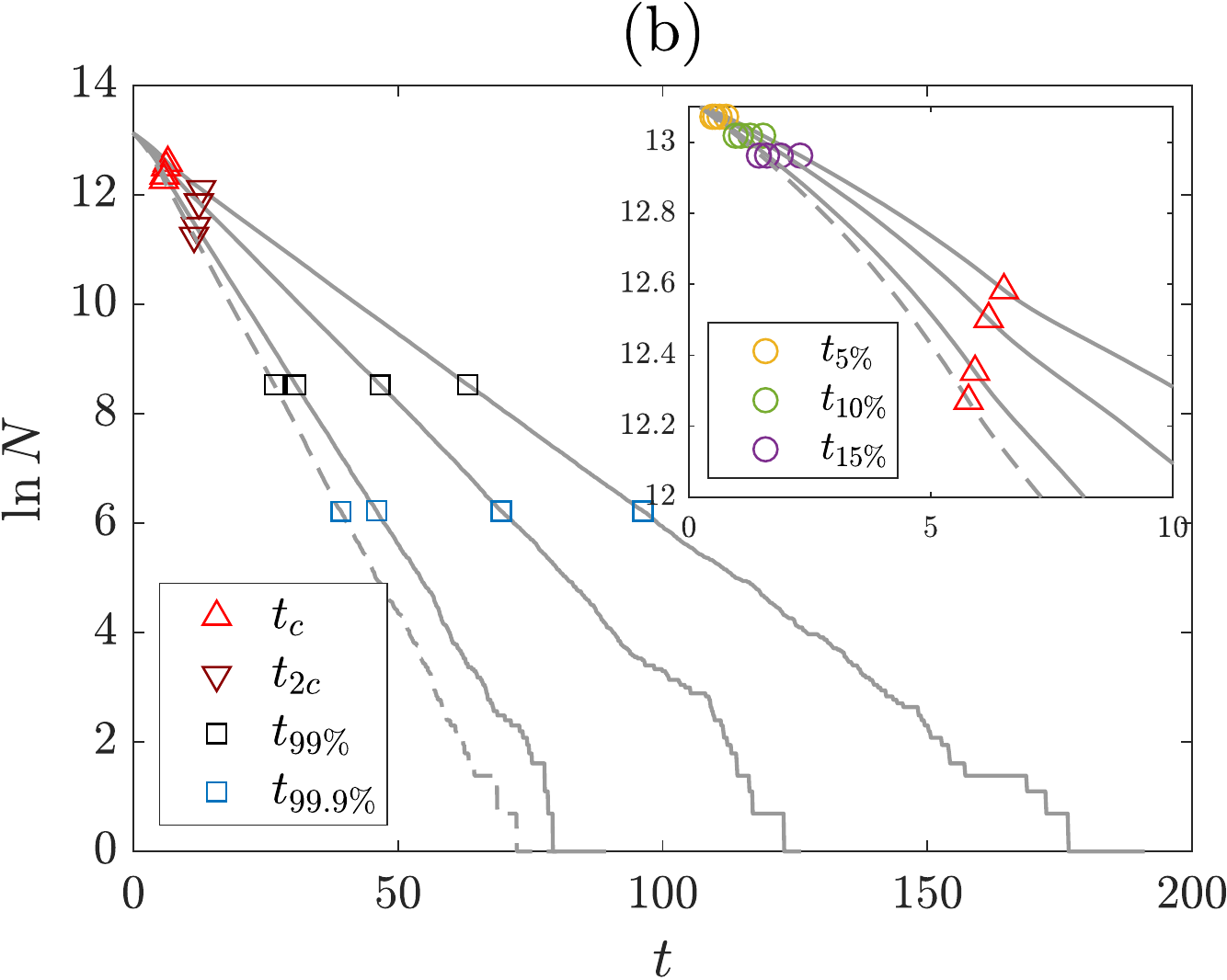}\\
	\bigskip
	\includegraphics[width=0.475\textwidth]{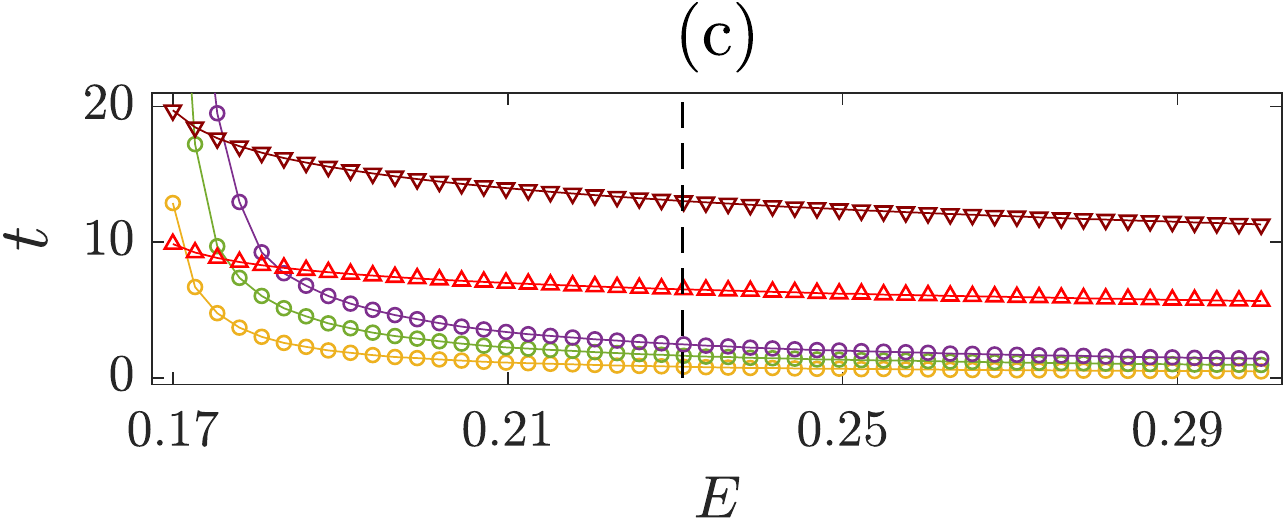}
	\quad
	\includegraphics[width=0.475\textwidth]{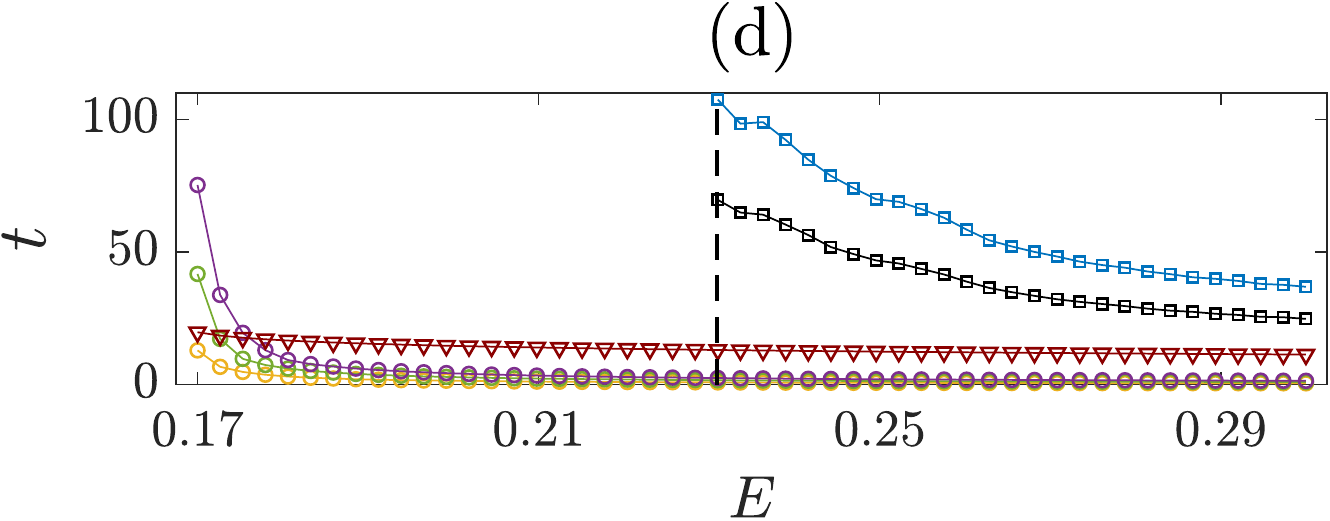}
	
	\caption{We consider short times ($t_{5\%}$, $t_{10\%}$ and $t_{15\%}$), the critical time ($t_c$) and long times ($t_{2c} \equiv 2t_c$, $t_{99\%}$ and $t_{99.9\%}$). (a) Decay curves (gray) when the underlying dynamics is nonhyperbolic for $E = 0.17$, $0.19$, $0.21$ (solid lines) and $0.23$ (dashed line). (b) Exponential decay curves (gray) when the underlying dynamics is hyperbolic, for $E = 0.235$, $0.25$, $0.275$ (solid lines) and $0.29$ (dashed line). (Inset plots) The first instants of the decay curves. (c) The evolution of the fitting times versus the system energy. (d) Again the evolution of the fitting times versus the system energy, but the evolution of the long times is included.}
	\label{fig:4}
\end{figure*}

To test the ergodic decay law, we launch a large amount of particles ($N_0 = 5 \cdot 10^5$) from different initial conditions randomly distributed in the configuration space of the scattering region, i.e., with random values for $x_0$, $y_0$ and $\theta_0$. For convenience, the evolution of the number of particles inside the potential well is shown in natural logarithmic scale, i.e., $\ln N(t)$ in Fig.~\ref{fig:4}. This representation enables the reader to infer easily an exponential decay when a linear relation between $\ln N$ and $t$ appears. We take into account several time intervals to fit the exponential behaviors.

In order to ascertain how the time intervals affects the decay, firstly, we define the short-time quantities $t_{5\%}$, $t_{10\%}$ and $t_{15\%}$, where, for instance, $t_{10\%}$ means the time at which the $10\%$ of the particles starting inside the scattering region have escaped. These short times are usually smaller than the critical time and thus are located at the beginning of the time evolution. Secondly, we consider the critical time, $t_c$. As indicated in the previous section, notice that the decay curve suffers a slight change of tendency when this time is reached in both dynamical regimes, although this change is even more subtle in the hyperbolic regime. Finally, we define the long-time quantity $t_{2c}$, which is twice the critical time for nonhyperbolic dynamics and, on the other hand, we also establish the long-time quantities $t_{2c}$, $t_{99\%}$ and $t_{99.9\%}$ for hyperbolic dynamics. Note that, in the nonhyperbolic regime, the exponential decay only appears during the first instants of the time evolution, provided that KAM tori turn the decay law algebraic at long times, as is clearly shown in Fig.~\ref{fig:4}(a). For this reason, we only compute exponential fittings until the time $t_{2c}$, when KAM tori are present. However, the presence of the chaotic saddle makes the escapes exponential in the hyperbolic regime (see Fig.~\ref{fig:4}(b)), and thus we are able to fit the exponential decay until longer times, such as $t_{99\%}$ or even $t_{99.9\%}$. Interestingly, we observe that the fraction of trapped particles for $E = 0.23$ is greater than for $E = 0.21$. This result is in agreement with the fact that the phase space volume occupied by KAM tori usually follows a non-trivial tendency as the energy changes. Furthermore, we show in Figs.~\ref{fig:4}(c) and \ref{fig:4}(d) how the time quantities defined above evolve as the system energy increases.

\begin{figure*}[htp!]
	\centering
	
	\includegraphics[width=0.475\textwidth]{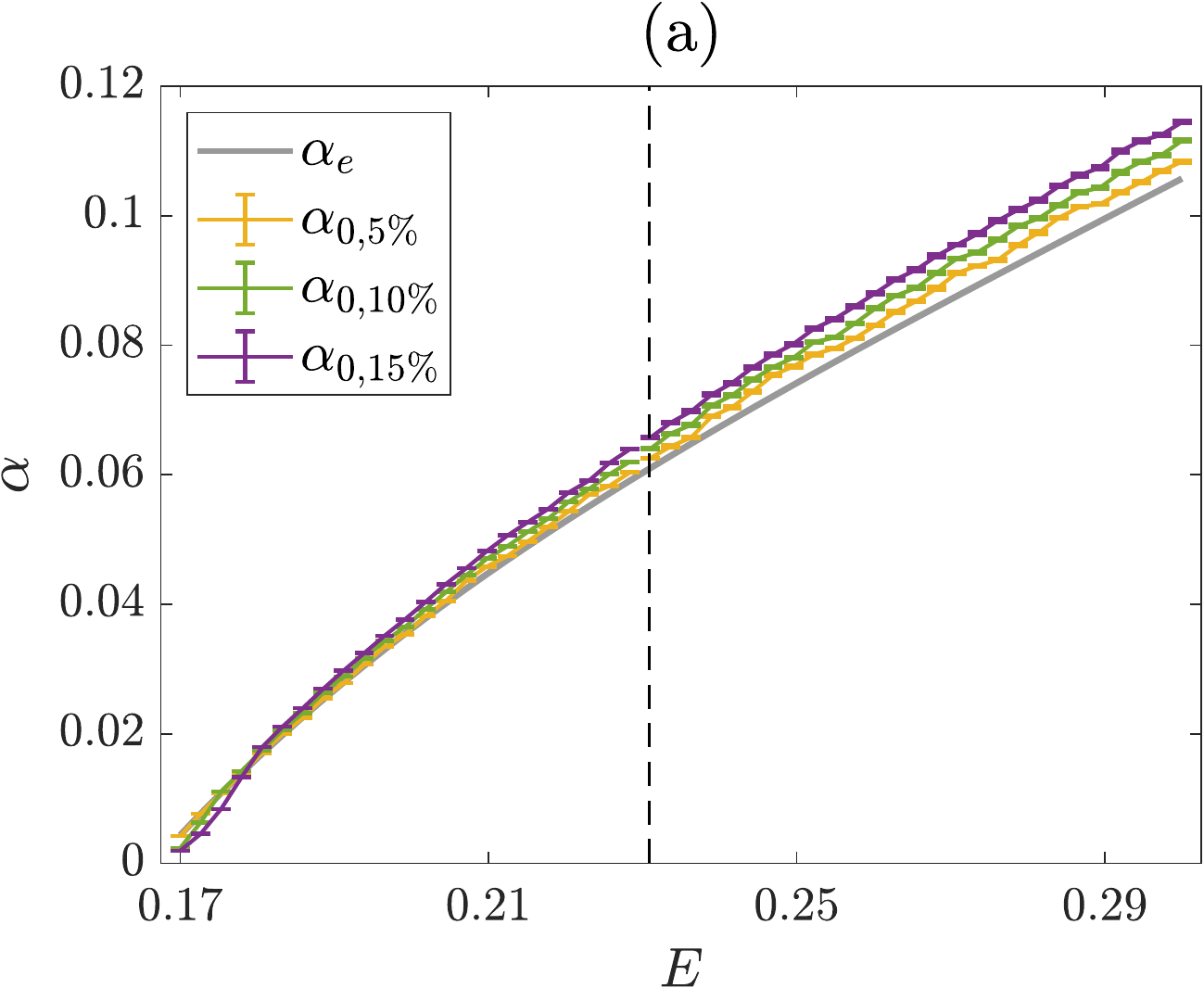}
	\quad
	\includegraphics[width=0.481\textwidth]{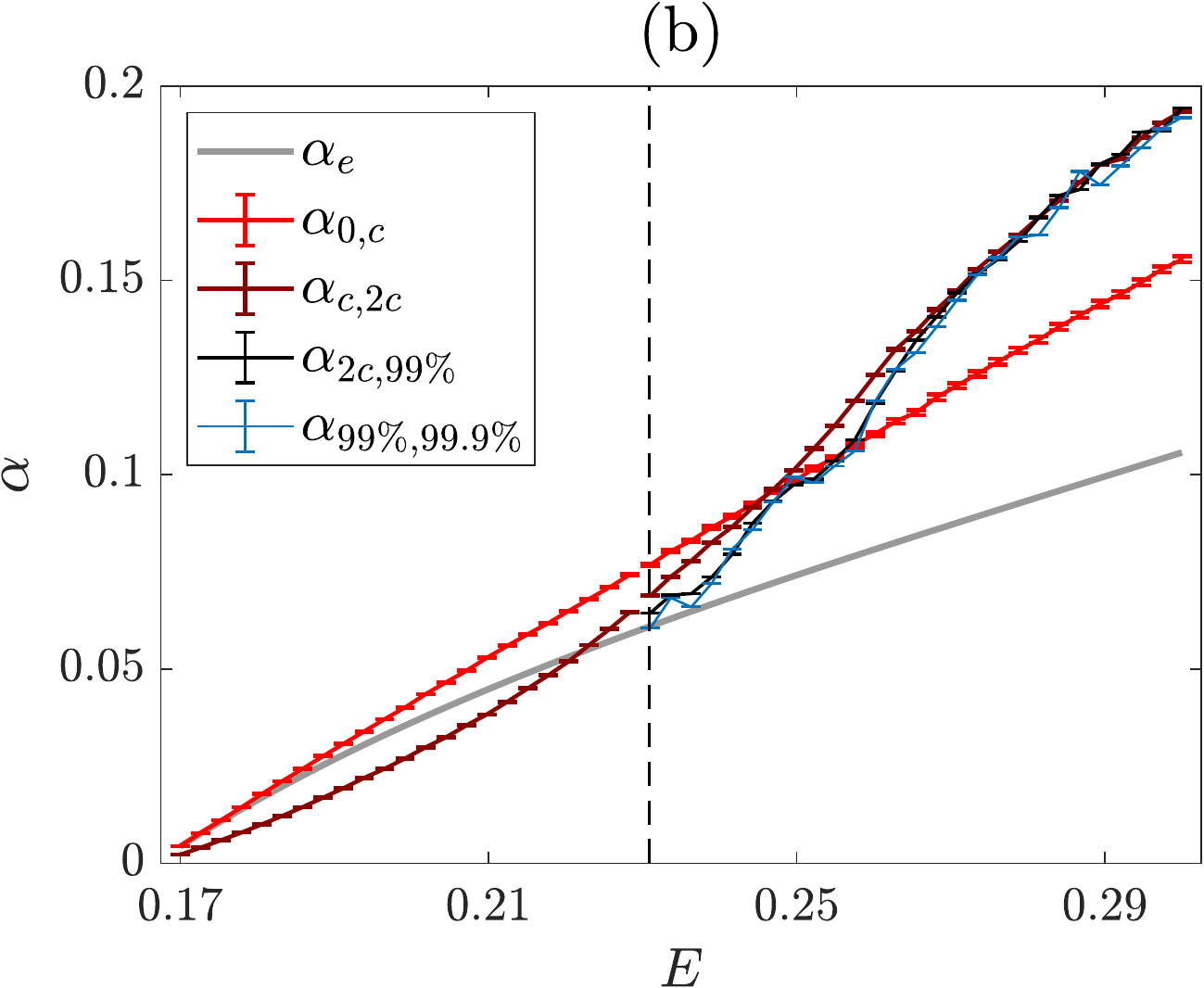}
	
	\caption{Exponential fittings of the decay curves at (a) short times, (b) critical and long times. We recall that the dashed line again splits the nonhyperbolic and hyperbolic dynamics when $E = 0.2309$. For the sake of clarity, although the decay law $\alpha_e$ has been computed in terms of $\Delta E$, we show the results of fittings in terms of $E$ for simplicity. 
	}
	\label{fig:5}
\end{figure*}

The fitted decay rates $\alpha_{a,b}$ associated with the time intervals $t \in [t_a,t_b]$ are displayed in Fig.~\ref{fig:5}. For example, the exponential rate $\alpha_{0,10\%}$ has been computed considering only a very small part of the decay curve, $t \in [0,t_{10\%}]$. We observe that the shorter the times considered in exponential fitting, the better the agreement is between the ergodic decay rate, $\alpha_e$, and the fitted rate, such as $\alpha_{0,5\%}$ (see Fig.~\ref{fig:5}(a)). This is explained because the ergodic decay law does not take into account the KAM tori and the chaotic saddle, whose traces in particle paths are more noticeable when the longer the time particles have evolved inside the scattering region. Thus, as a consequence, the ergodic decay law overestimates the decay rate $\alpha_{c,2c}$ when KAM tori are present and escapes are delayed by them. However, in the hyperbolic regime, the ergodic decay law underestimates all long-time decay rates, such as $\alpha_{c,2c}$, $\alpha_{2c,99\%}$ and $\alpha_{99\%,99.9\%}$, because the unstable manifold of the chaotic saddle affects the escapes at long times, accelerating them (see Fig.~\ref{fig:5}(b)). In addition, we note that as the system energy is increased, the chaotic saddle also enhances its effects on escapes at short times. Hence, the ergodic decay rate underestimates escapes when the energy system is high even at short times. This last argument can be applied when analyzing why the fitted rate $\alpha_{0,c}$ does not agree with the ergodic decay law neither, because the influence of the sets that governs the dynamics is strong enough to prevent particles from being distributed according to the microcanonical ensemble at the critical time as well.

Furthermore, we would like to highlight that the decay rates for long times may be useful to quantify whether particles always escape in the same way over the three different periods of the time evolution that we establish in the hyperbolic regime, namely $\alpha_{c,2c}$, $\alpha_{2c,99\%}$ and $\alpha_{99\%,99.9\%} $. For example, these decay rates can be affected by the existence of extremely small KAM tori that may survive even after the destruction of the dominant KAM tori \cite{barrio2009,barrio2020}. In our simulations, we have not found trapped particles for energies greater than $ E> 0.2309 $, neither long-time algebraic decays for at least the $99.9\%$ of the initialized particles in the scattering region. Therefore, such remnant KAM structures are not decisive in the exponential escape regime. Nonetheless, we observe in Fig.~\ref{fig:5}(b) that the decay rates of the two last periods considered, $\alpha_{2c,99\%}$ and $\alpha_{99\%,99.9\%}$, exhibit a fluctuating evolution when plotted against the system energy, whose cause deserves further research. So far, we have found that hyperbolic decay rates measured at any given sufficiently long time are underestimated by the ergodic decay law, which implies a clear limitation of this kind of decay laws in describing escapes in open systems.

\begin{figure*}[htp!]
	\centering
	
	\includegraphics[width=0.3125\textwidth]{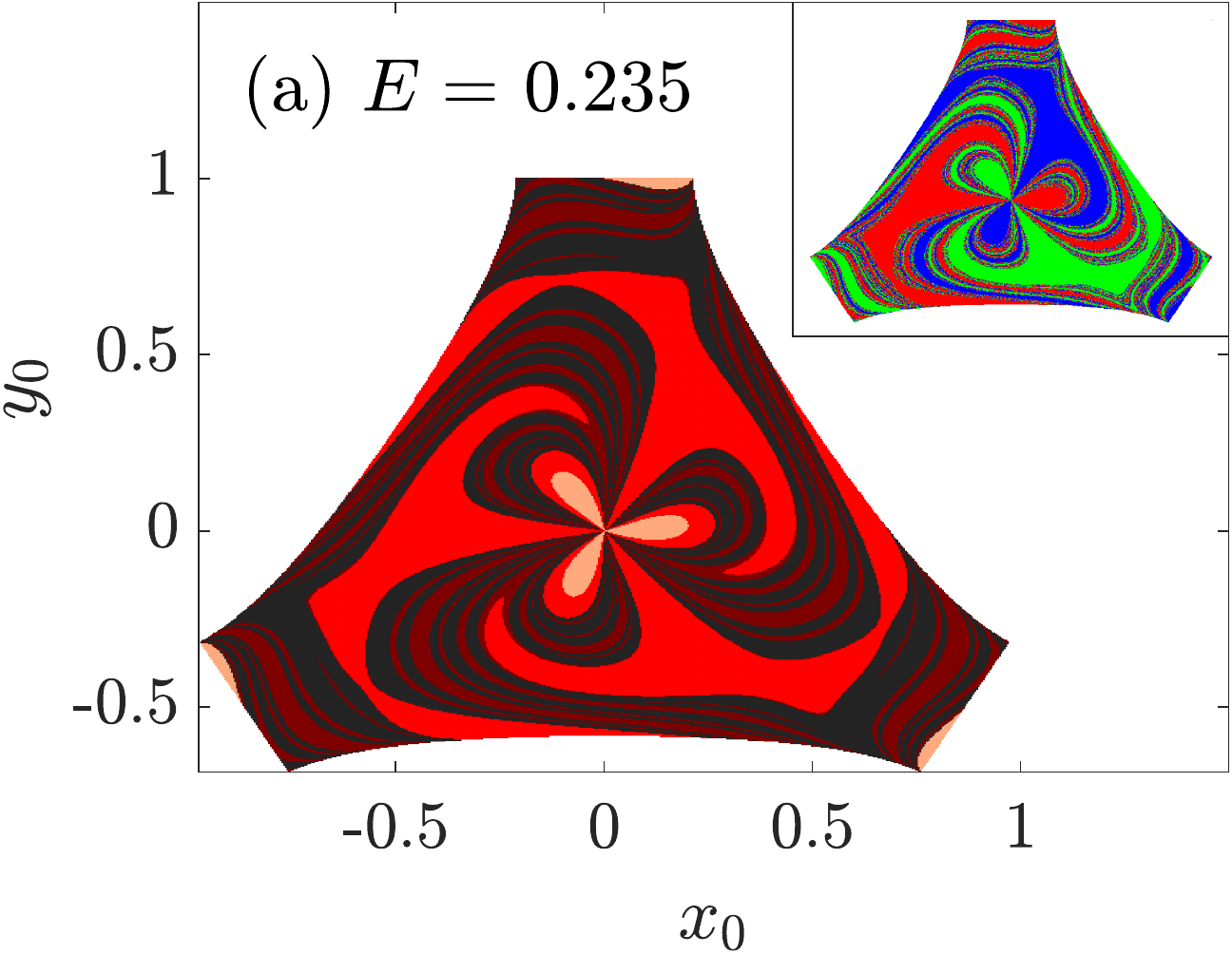}
	\quad
	\includegraphics[width=0.3125\textwidth]{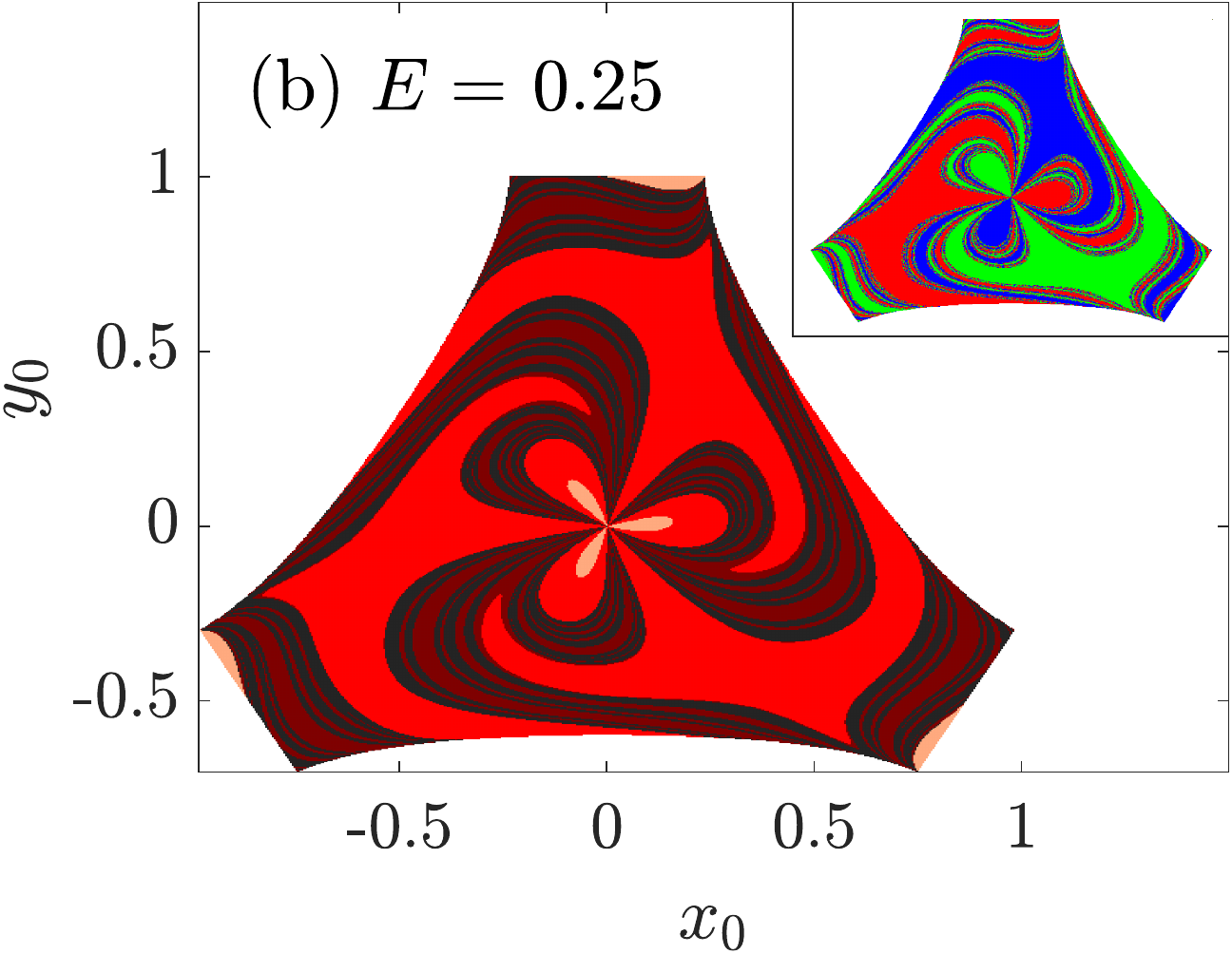}
	\quad
	\includegraphics[width=0.3125\textwidth]{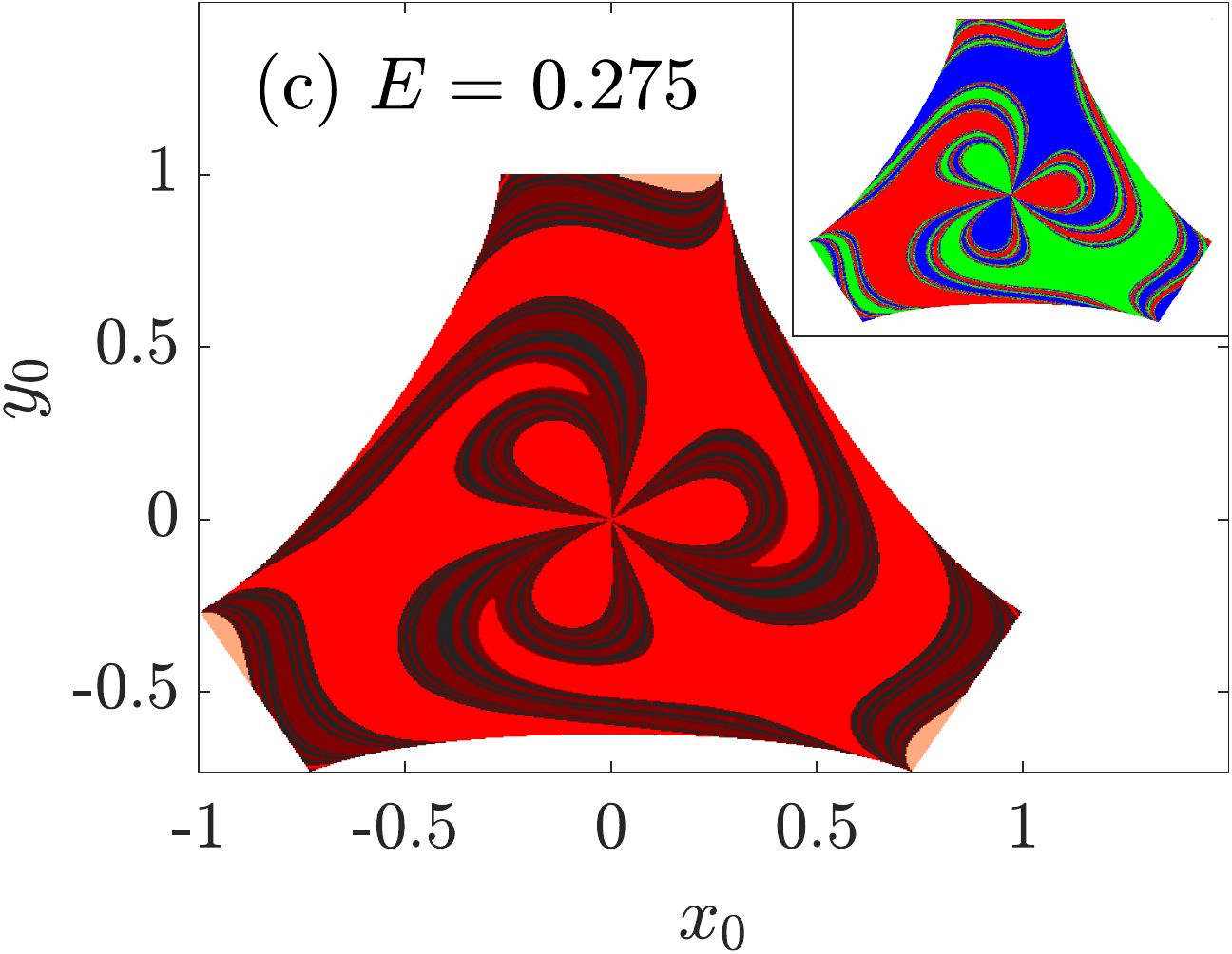}\\
	\bigskip
	\includegraphics[width=0.3125\textwidth]{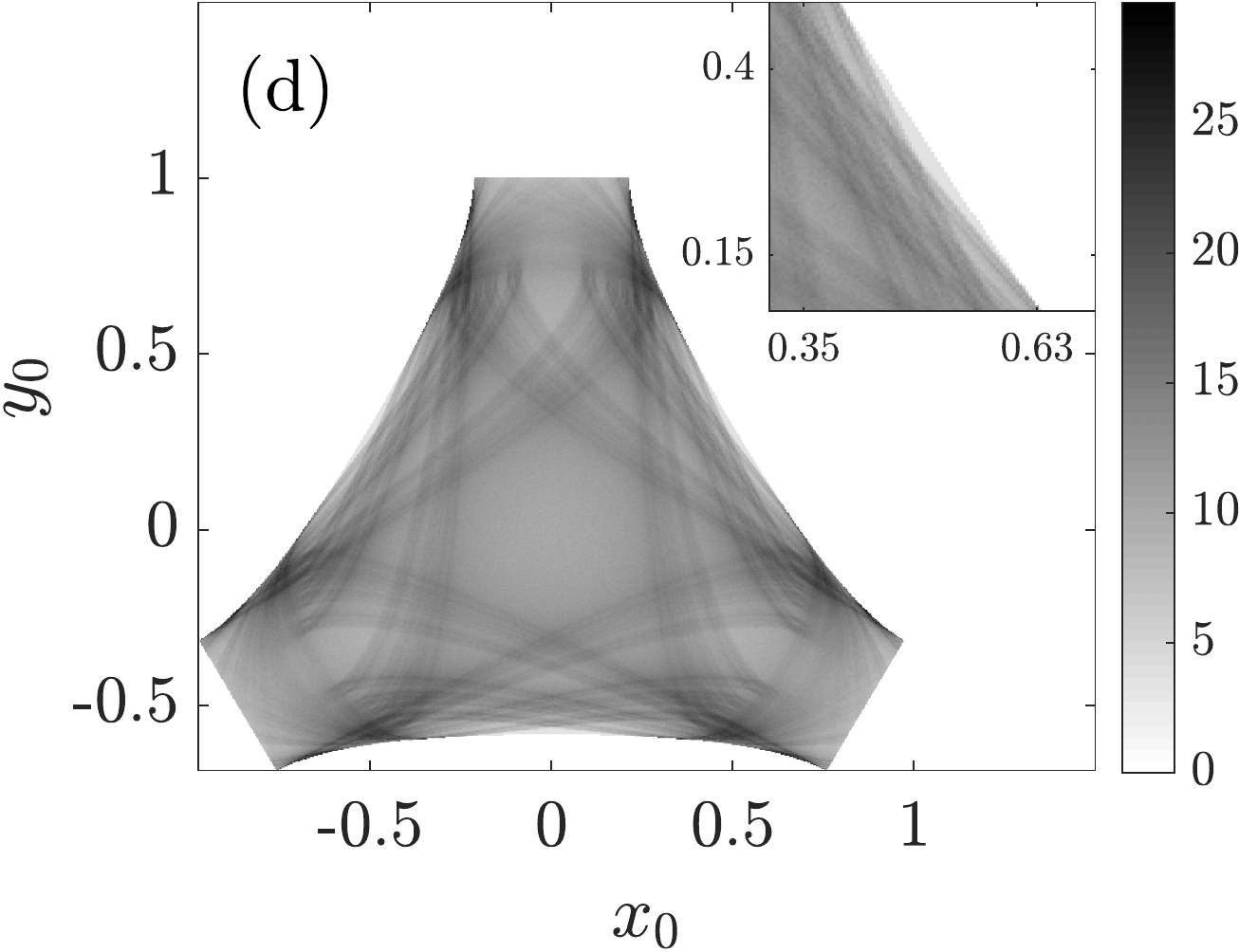}
	\quad
	\includegraphics[width=0.3125\textwidth]{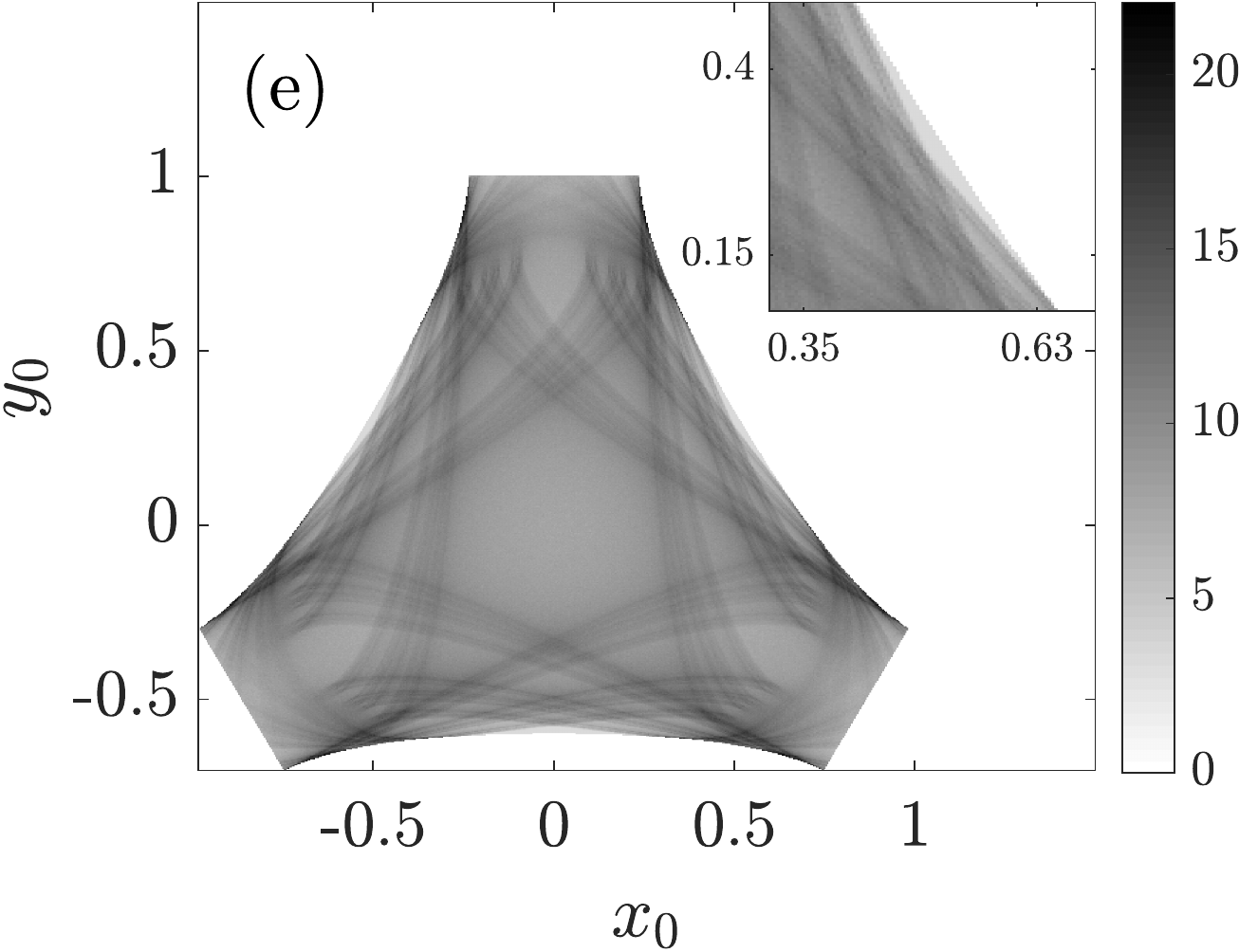}
	\quad
	\includegraphics[width=0.3125\textwidth]{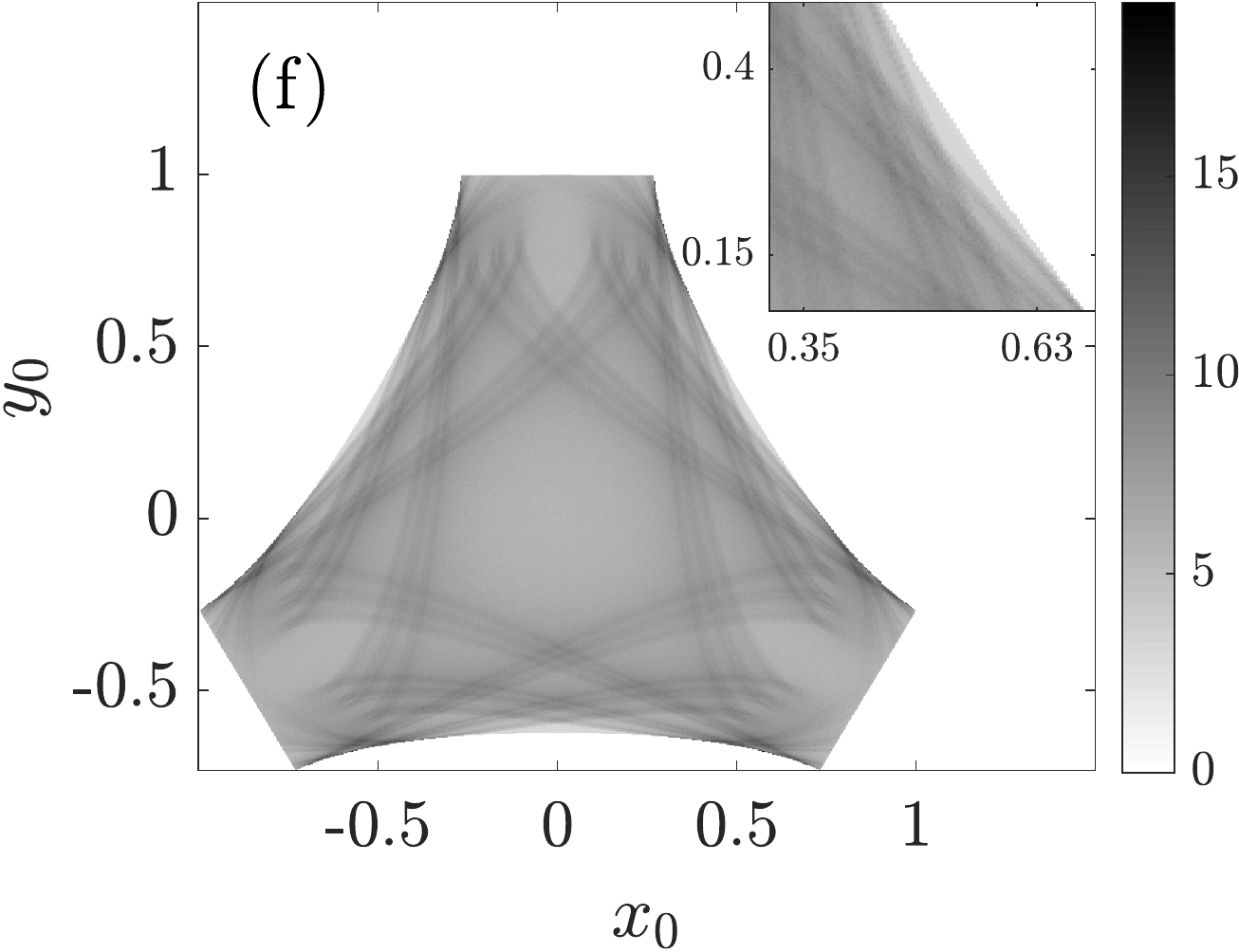}
	
	\caption{(a--c) Decay basins for $E = 0.235$, $0.25$ and $0.275$, respectively. The color code indicates the initial conditions from which escapes occur before the short time $t_{15\%}$ (lighter red), the critical time $t_c$ (red), twice the critical time $t_{2c}$ (darker red), and finally those from which particles have not already escaped by $t_{2c}$ (black). (Inset plot) Exit basins associated with the shown decay basins. (d--f) Average escape times, again, for $E = 0.235$, $0.25$ and $0.275$. We launch $2.5 \cdot 10^3$ particles for each initial condition $(x_0,y_0)$ with random shooting angles belonging to the interval $\theta_0 \in [0,2\pi)$. As explained in the text, the average escape times can be interpreted as the regions where it is more likely to find a particle in the configuration space. (Inset plot) Zoom-in near the potential barrier between Exit 1 and Exit 3, where a low-probability region is found in the three cases.}
	\label{fig:6}
\end{figure*}

Now, we present the \emph{decay basins} for hyperbolic dynamics in Figs.~\ref{fig:6}(a--c) to visualize which particles escape during each time interval and how they are distributed inside the scattering region. Furthermore, they relate the escape times and the exit basins built by utilizing the tangential shooting method \cite{aguirre2001}. We observe that an important fraction of particles escapes before the critical time. These particles start their escaping trajectories far from the fractal boundaries of the exit basins, i.e., far from the influence of the chaotic saddle. This confirms that the critical time is the maximum escape time for a trajectory that is not affected by the sets that governs the dynamics. Conversely, all the initial conditions associated with escape times higher than the critical time are located close to the fractal boundaries of the exit basins.

Nonetheless, we are interested in the escape time distribution by shooting particles from initial conditions randomly. Hence, we show here the average escape times in the configuration space in Figs.~\ref{fig:6}(d--f), where for each pixel we launch a large amount of particles and compute the mean of their escape times. In the three cases shown, it can be observed an apparently fractal structure that is reminiscent of the typical structures that appear in scattering functions when a Smale horseshoe-like set rules the dynamics, such as the chaotic saddle. In fact, this structure is formed by the initial conditions that spend long escape times, i.e., the initial conditions close to the stable manifold of the chaotic saddle \cite{ott1993_2}. Therefore, the images of average escape times provide a clear manifestation of the presence of the chaotic saddle when particles are launched randomly.

Moreover, we study how the chaotic saddle explicitly affects particle trajectories. Then, again, by launching particles with random values for $x_0$, $y_0$ and $\theta_0$, and taking into account every position of their orbits, we obtain the regions where a hypothetical particle is more likely to be found before its eventual escape. Not surprisingly, the images obtained are the same as those provided by calculating the average escape times (see Figs.~\ref{fig:6}(d--f)), since trajectories starting close to the chaotic saddle spend longer times within the potential well: they come close to the chaotic saddle along its stable manifold, stay in its neighborhood, and leave it along its unstable manifold. In light of the results, the particle is more likely to be found near the chaotic saddle in its escaping process. In addition, we also detect a region of low probability of finding the particle, as can be visualized in the inset plots of Figs.~\ref{fig:6}(d--f).

\begin{figure*}[htp!]
	\centering
	
	\includegraphics[width=0.3125\textwidth]{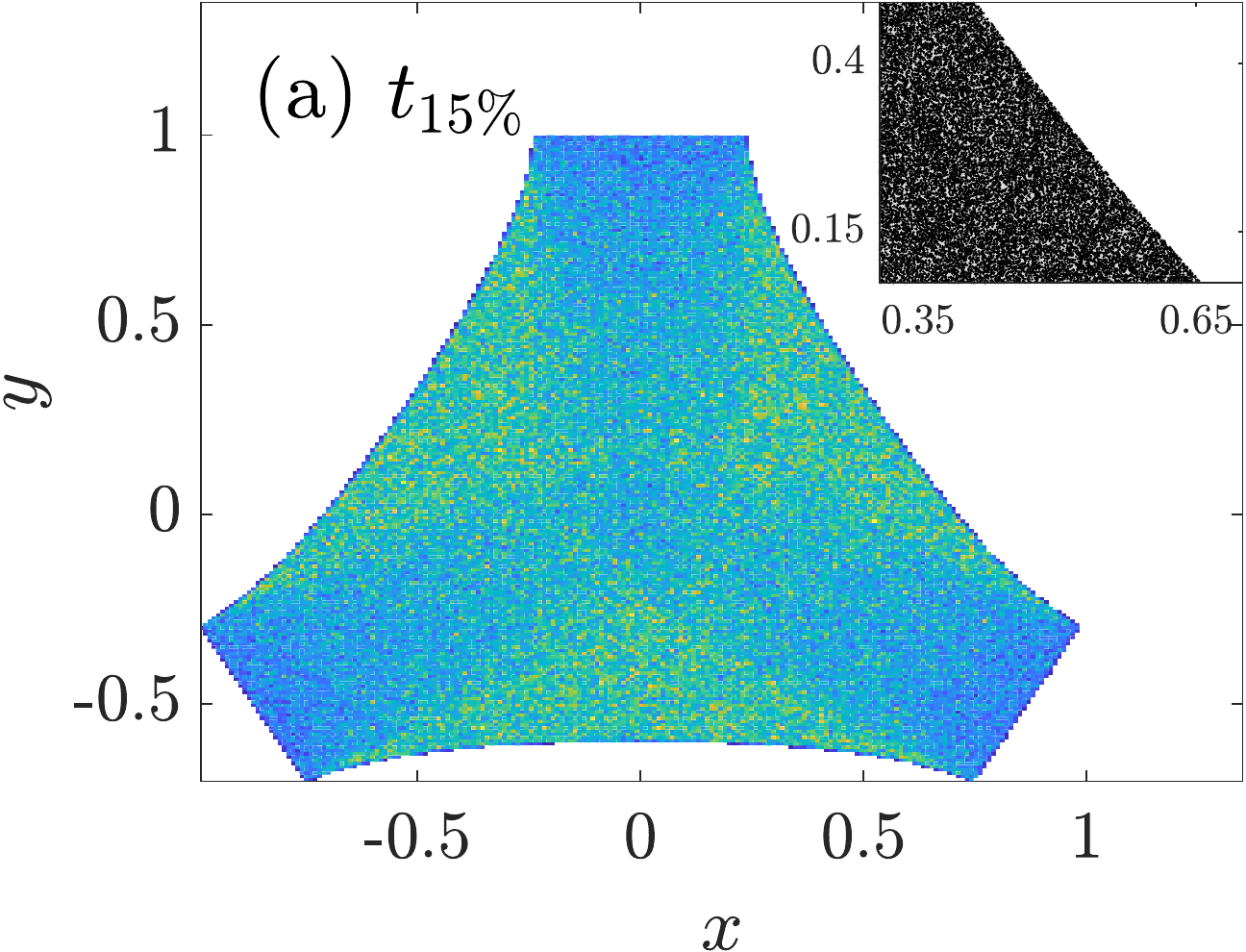}
	\quad
	\includegraphics[width=0.3125\textwidth]{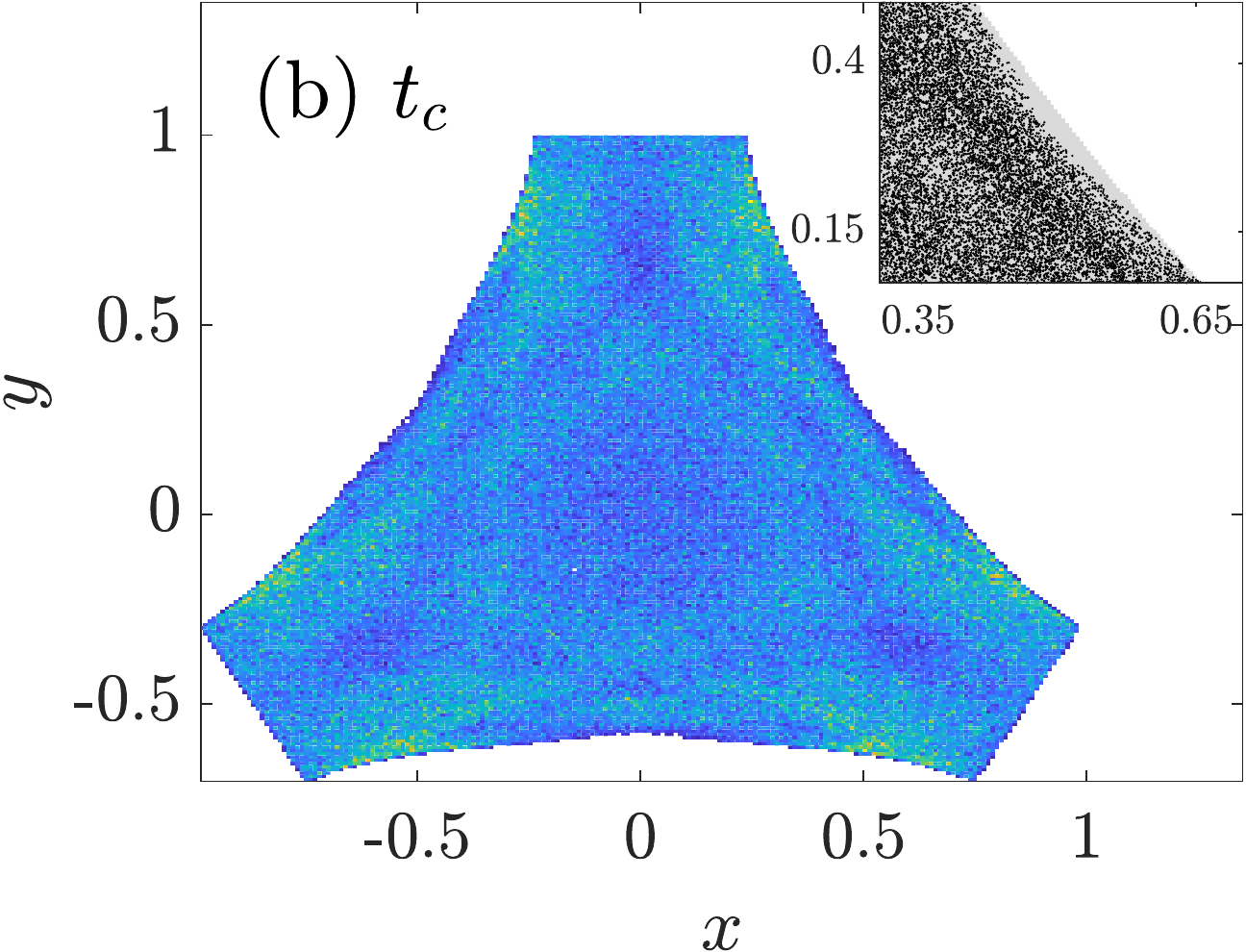}
	\quad
	\includegraphics[width=0.3125\textwidth]{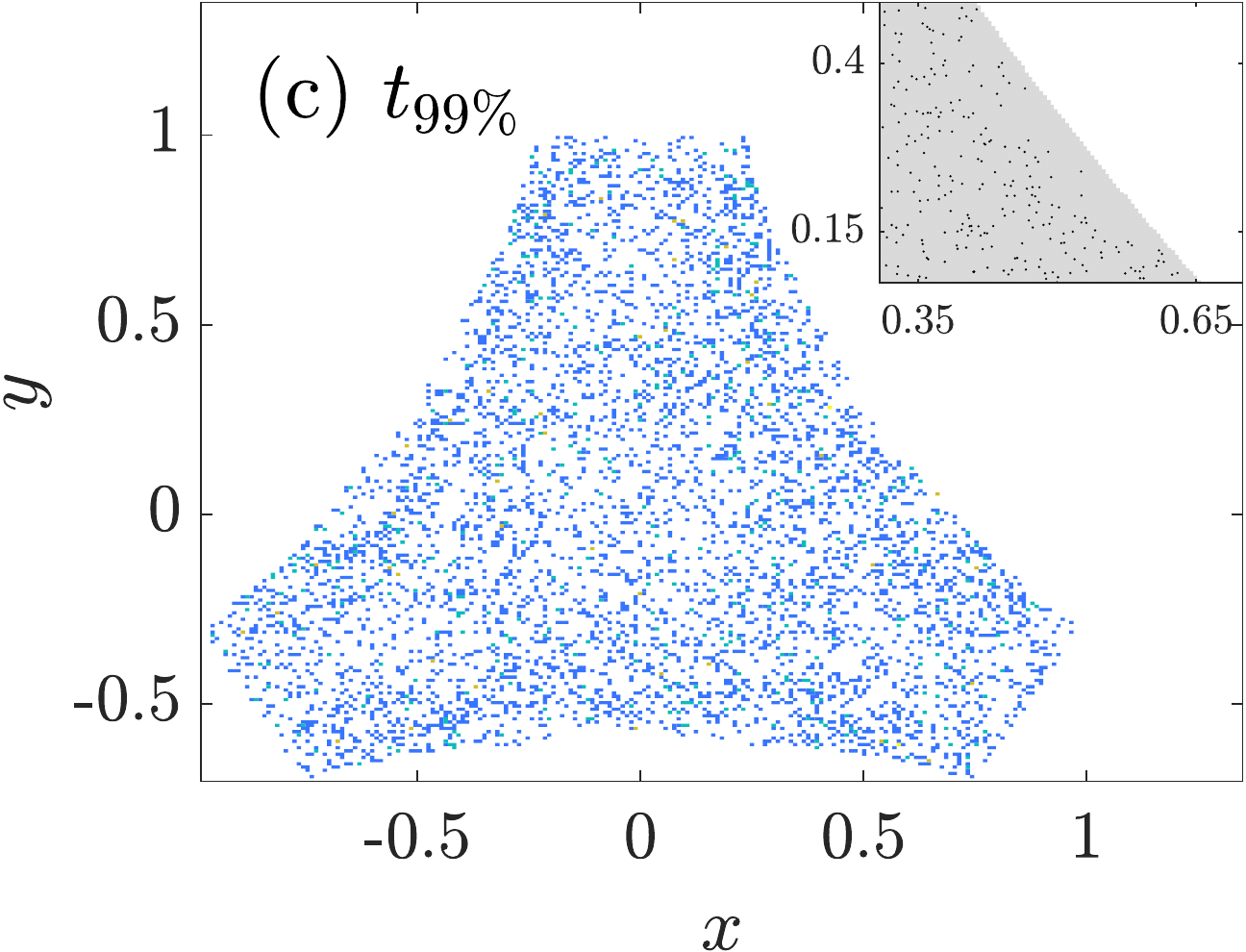}
	
	\caption{Particle distribution at (a) $t_{15\%}$, (b) $t_c$ and (c) $t_{99\%}$, respectively, by means of a histogram, for $E = 0.25$. The histogram color code consists of a fading between yellow (highly occupied area) and dark blue (lowly occupied area). The white color means an area depleted of particles. (Inset plots) Zoom-in of the particle distribution near the potential barrier between Exit 1 and Exit 3. We depict the present particles (black dots) in the allowed scattering region compatible with energy (gray).}
	\label{fig:7}
\end{figure*}

Finally, in Fig.~\ref{fig:7} we show the instantaneous distribution of particles at some relevant times of the decay law. The regions that are emptied in the first place are those close to the exits from which particles can escape if the shooting angle points directly towards the exit. Subsequently, when the critical time is reached, a symmetric structure can be inferred in the particle distribution, which is closely related to the manifestation of the chaotic saddle (see Figs.~\ref{fig:6}(d--f)). Interestingly, the fact that trajectories spend a finite period of time in the chaotic saddle vicinity implies that there are zones of the scattering region full of particles and others empty during the time evolution. For instance, as indicated above, there is an almost zero probability of staying in the space nearby the wall between two exits, because the chaotic saddle does not occupy that zone (see the inset plots of Figs.~\ref{fig:7}(a--b)). Hence, the presence of the chaotic saddle is incompatible with an equiprobability of being in any dynamical state compatible with energy, as assumed by the microcanonical ensemble.

\section{Decay laws in the relativistic H\'{e}non-Heiles system} \label{sec:4}

We now employ a relativistic version of the H\'{e}non-Heiles system to extent our claims to the relativistic realm. Hence, we consider a particle interacting in the limit of weak external fields whose dynamics is governed by the conservative and dimensionless Hamiltonian \cite{lan2011,chanda2018,kovacs2011,calura1997} \begin{equation} H(p_x, p_y, x, y) = c \sqrt{c^2 + p_x^2 + p_y^2} + V(x,y), \label{eq:7} \end{equation} where $c$ is the value of the speed of light and $V(x,y)$ is again the H\'{e}non-Heiles potential described above. Although the equations of motion are different from the Newtonian ones, the four fixed points of the relativistic system still remain at the three saddle points of the potential well and at its minimum. For this reason, we have defined the relativistic scattering region as in the Newtonian H\'{e}non-Heiles system. Furthermore, the Lorentz factor is defined as \begin{equation} \gamma = \frac{1}{\sqrt{1 - \frac{v^2}{c^2}}} = \frac{1}{\sqrt{1 - \beta^2}}, \label{eq:8} \end{equation} where $v$ is the speed of the particle and $\beta = v/c$ the ratio between the speed of the particle and the speed of light. The Lorentz factor $\gamma$ and the quotient $\beta$ are two ways to express how large is $v$ compared to $c$, and they vary in the ranges $\gamma \in [1,+\infty)$ and $\beta \in [0,1)$, respectively. We shall use $\beta$ as the relativistic parameter for convenience in showing our numerical results.

We utilize a method introduced in \cite{bernal2017, bernal2018} to increase the kinetic energy of the system to the relativistic regime (the value of $\beta$). Firstly, the maximum speed that the particle can have along its trajectory before escaping can be defined as $v_m \equiv \sqrt{2E}$, bearing in mind that the maximum kinetic energy is found at the potential minimum, $V(x_m,y_m) = 0$, and $E$ is the total mechanical energy of the Newtonian version of the H\'{e}non-Heiles. Thus, the method is based on the fact that $\beta$ is a quantity that equivalently depends on $v_m$ and $c$, hence working on a dimensionless system it is possible to increase the value of $\beta$ by fixing the value of $v_m$ and decreasing the numerical value of the speed of light. In this manner, if the value of $\beta$ is almost null, $v_m \ll c$, the speeds of the particle during its evolution within the scattering region only represents a very low percentage of the speed of light. In this case, we recover the Newtonian version of the H\'{e}non-Heiles system. On the contrary, if the value of $\beta$ is close to one, the speed of the particle represents a high percentage of the speed of light. For the sake of clarity, although the numerical value of $c$ is modified to select a value of $\beta$, this numerical value of $c$ always represents the speed of light in the system because its dynamics is governed by the Hamiltonian \eqref{eq:7}. For more method's details see \cite{fernandez2020}.

\begin{figure*}[htp!]
	\centering
	
	\includegraphics[width=0.3125\textwidth]{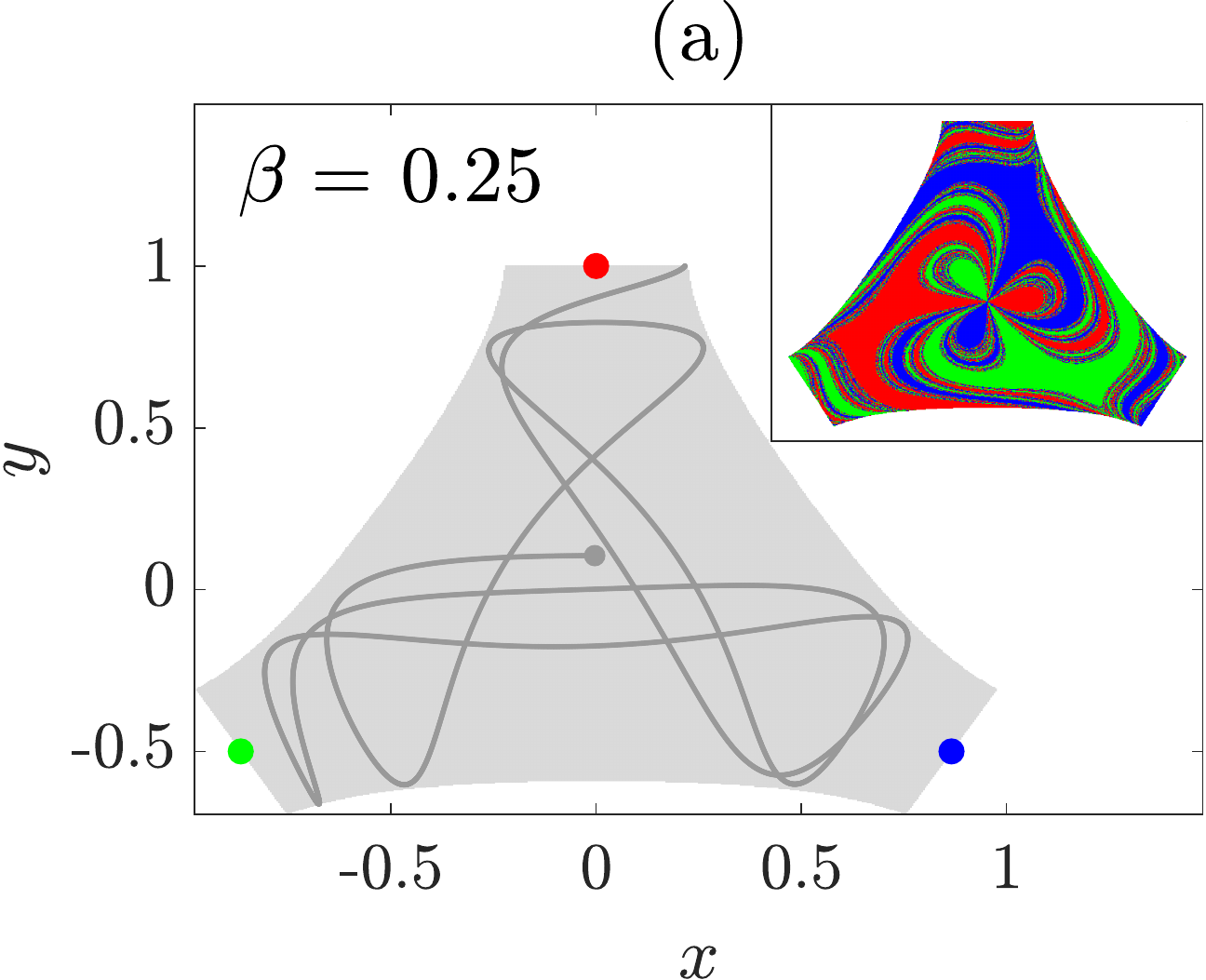}
	\quad
	\includegraphics[width=0.3125\textwidth]{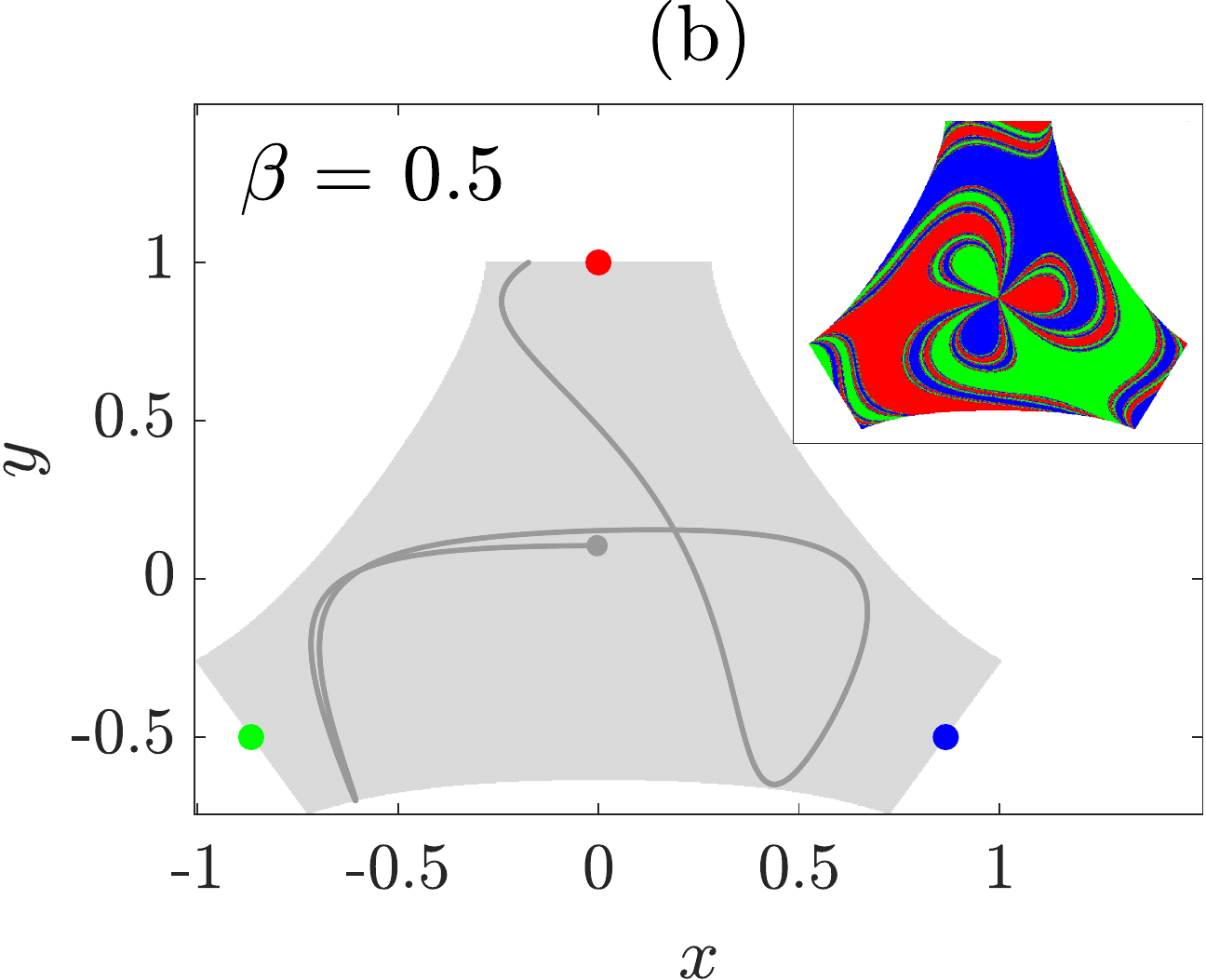}
	\quad
	\includegraphics[width=0.3125\textwidth]{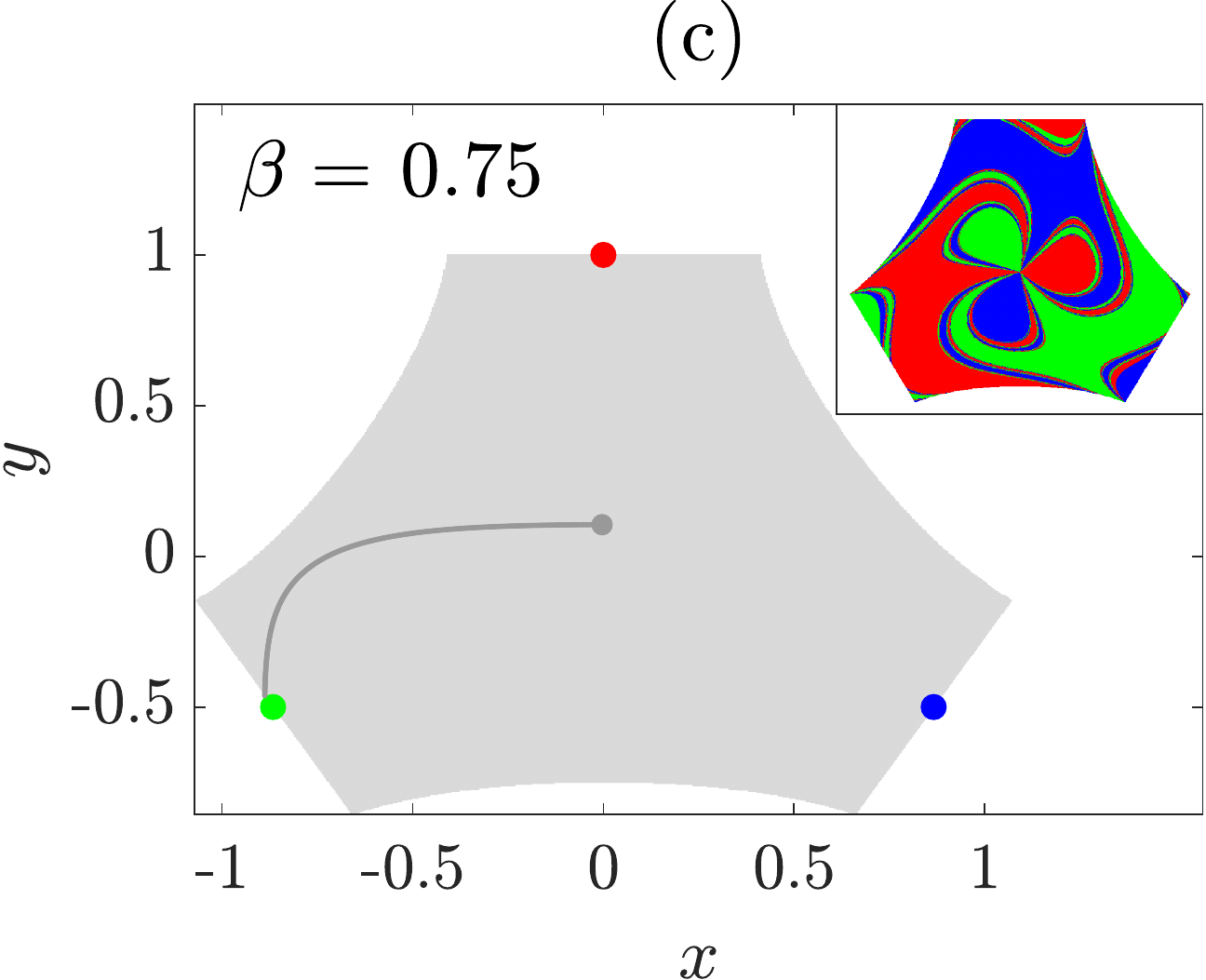}
	
	\caption{Scattering regions (pale gray) and escaping trajectories starting from the same condition but for (a) $\beta = 0.25$, (b) $0.5$ and (c) $0.75$, where the underlying dynamics is hyperbolic. As $\beta = v_m/c$ always holds, the computational value of the speed of light changes inevitably when the value of $\beta$ changes. We have arbitrarily set $v_m \approx 0.679$, and therefore the values of the speed of light in the computations are (a) $c \approx 2.718$, (b) $1.359$ and (c) $0.906$, respectively. (Inset plots) Exit basins associated.}
	\label{fig:8}
\end{figure*}

We have arbitrarily established $v_m \approx 0.679$ along this section, which corresponds to the Newtonian energy $E \approx 0.2309$, to focus on hyperbolic dynamics in this relativistic H\'{e}non-Heiles system. Summarizing, the Newtonian and hyperbolic system becomes relativistic and increasingly hyperbolic as $\beta$ increases. In Fig.~\ref{fig:8}, we show the effect of $\beta$ on the dynamics and the basin topology: the particle generally escapes more quickly and the fractal boundaries of exit basins are smoother as $\beta$ increases. Finally, the computations along this section have been carried out only until $\beta = 0.75$, since the relativistic Hamiltonian \eqref{eq:7} has been derived in the limit of weak external fields.

We also concentrate our efforts on exponential decay laws along this section and derive a decay rate based on ergodic motion inside the scattering region. Firstly, instead of the momenta $p_x$ and $p_y$, Eq.~\eqref{eq:7} can be rewritten in terms of the total momentum of the particle $p$ and the angle $\theta$ formed by the speed vector and the positive $x$-axis, obtaining thus \begin{equation} H(p,x,y,\theta) = c \sqrt{c^2 + p^2} + V(x,y). \label{eq:9}\end{equation} Importantly, we recall that the total momentum of the particle is related to its speed in this dimensionless version of the system by $p \equiv \gamma v$. Furthermore, we express the total relativistic energy as $\Delta E \equiv E - E_e = \gamma c^2 + V - 1/6$ for convenience when computing the decay rate. In addition, there exists a relation between $\Delta E$ and $\beta$, which is \begin{equation} \Delta E(\beta) = \frac{c^2}{\sqrt{1-\beta^2}} - \frac{1}{6}, \label{eq:10}\end{equation} since the potential energy is null, $V(x_m,y_m) = 0$, when we define $\beta$ as the relativistic parameter of the system. The motion in the well is assumed to be ergodic, and therefore the phase space distribution inside the scattering region for a given energy $\Delta E$ is expressed by \begin{equation} \rho(p,x,y,\theta) = \frac{\delta \left( E - H(p,x,y,\theta)\right)}{ \int dp dx dy d\theta \delta \left( E - H(p,x,y,\theta)\right)}, \label{eq:11}\end{equation} where $\delta(E)$ represents the Dirac delta function. Here, we are interested in the distribution of the variables in the configuration space. In this manner, we integrate Eq.~\eqref{eq:11} along the interval $p \in [0,\infty)$, yielding \begin{equation} \rho(x,y,\theta) = \frac{\gamma}{2\pi \int_{S(\Delta E)} dx dy \gamma}, \label{eq:12} \end{equation} where the integral is bounded inside the scattering region, and the Lorentz factor $\gamma$ depends on the spatial coordinates because the system is conservative and every point of the well is associated with a specific kinetic energy value. Notice that if the limit $c \to \infty$ (or $\gamma \to 1$, equivalently) is considered, the Newtonian probability density $\rho(x,y,\theta) = 1/2\pi S(\Delta E)$ is recovered, as derived in \cite{zhao2007}. For simplicity, we define $\Gamma (\Delta E) \equiv \int_{S(\Delta E)} dxdy\gamma$, where the integral can be solved numerically by means of a Monte Carlo method.

\begin{figure*}[b!]
	\centering
	
	\includegraphics[width=0.475\textwidth]{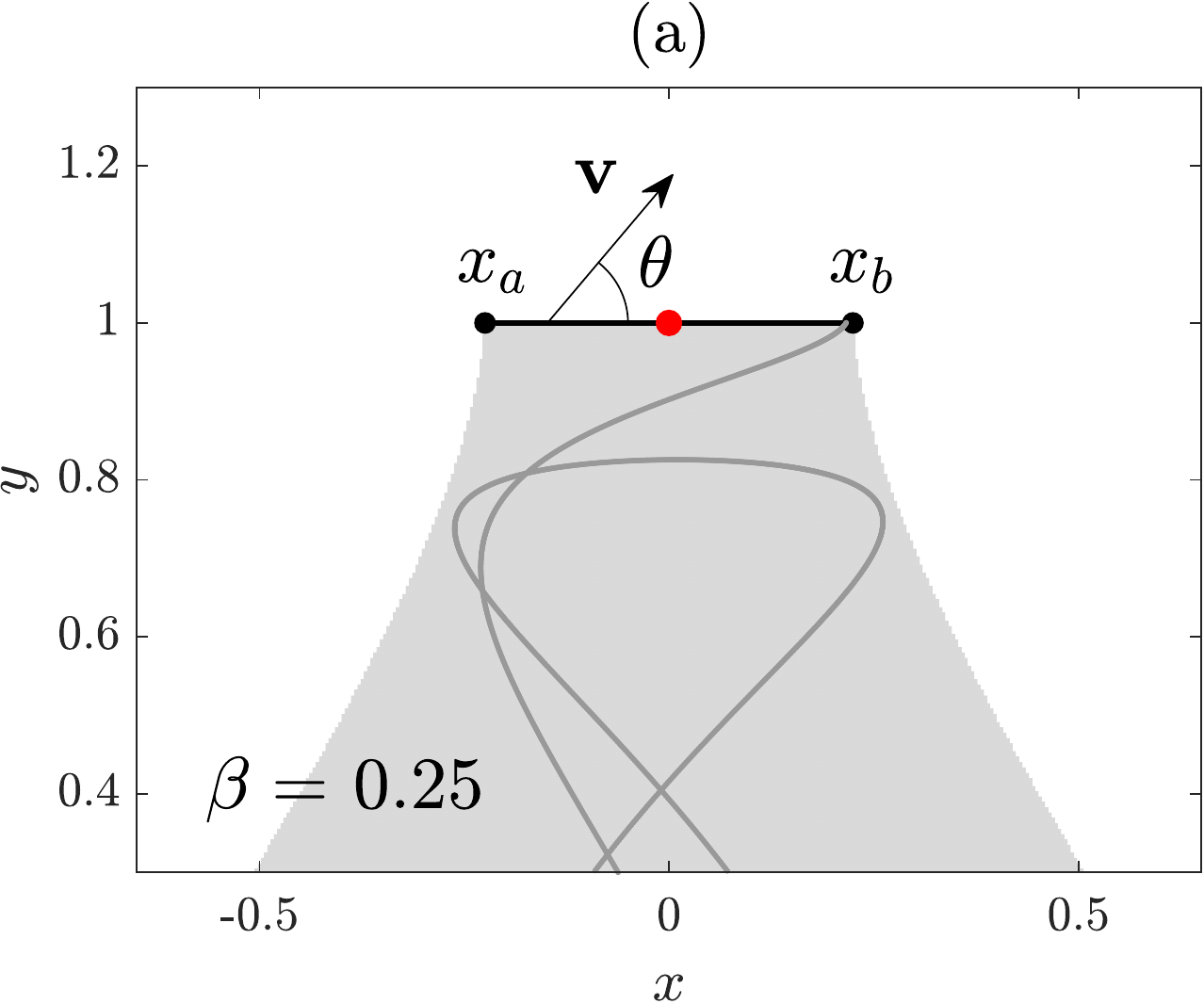}
	\quad
	\includegraphics[width=0.475\textwidth]{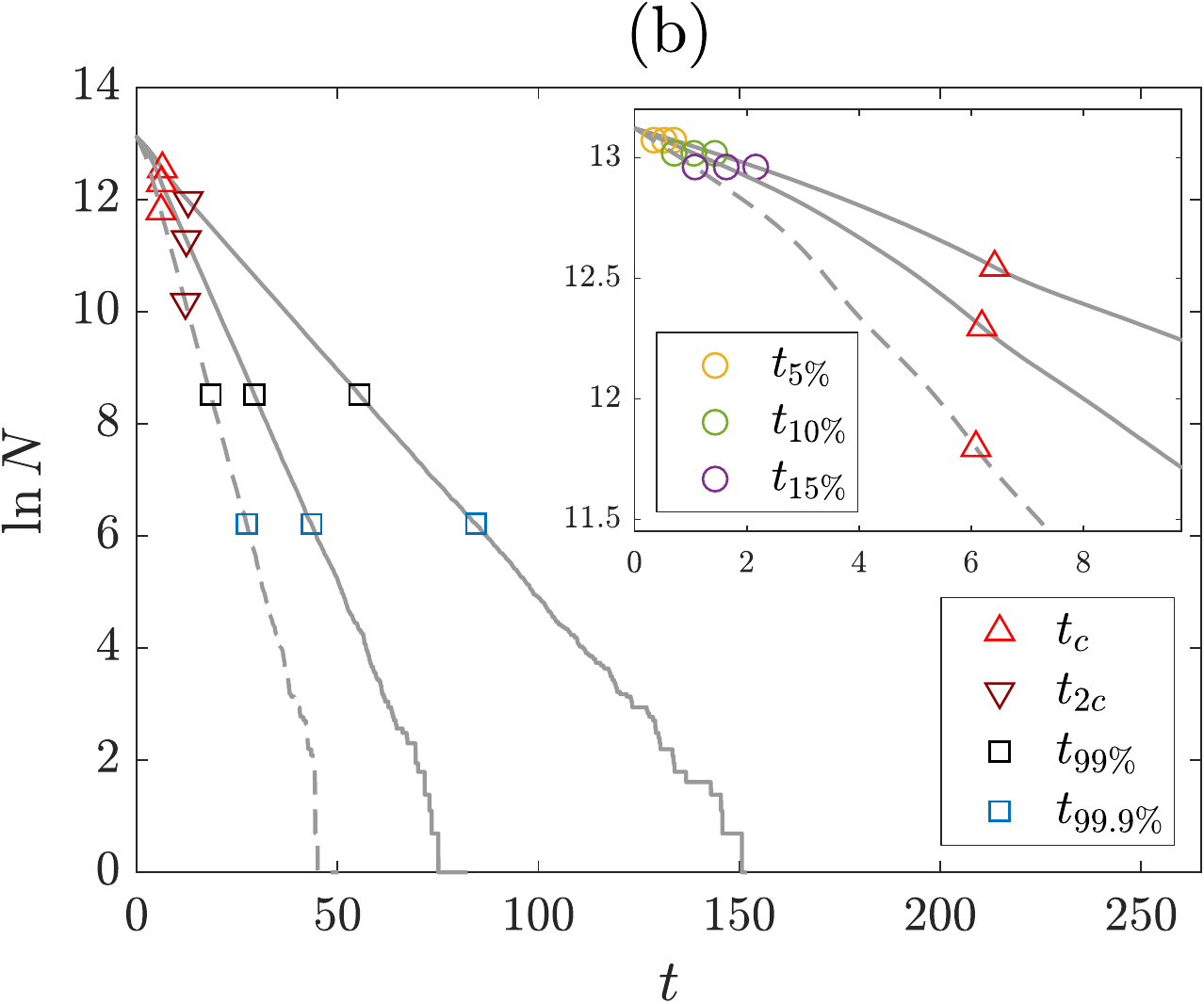}\\
	\bigskip
	\includegraphics[width=0.475\textwidth]{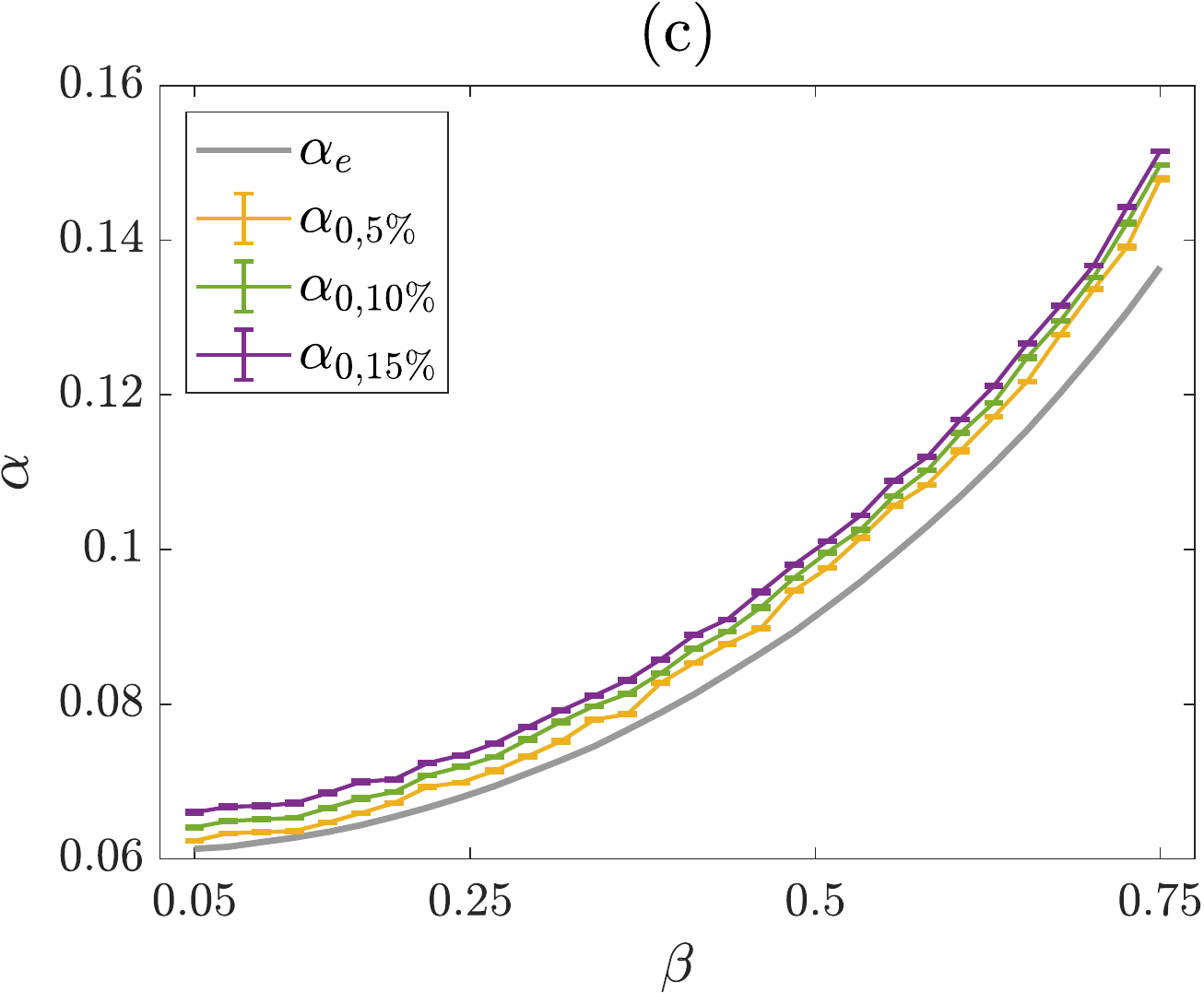}
	\quad
	\includegraphics[width=0.485\textwidth]{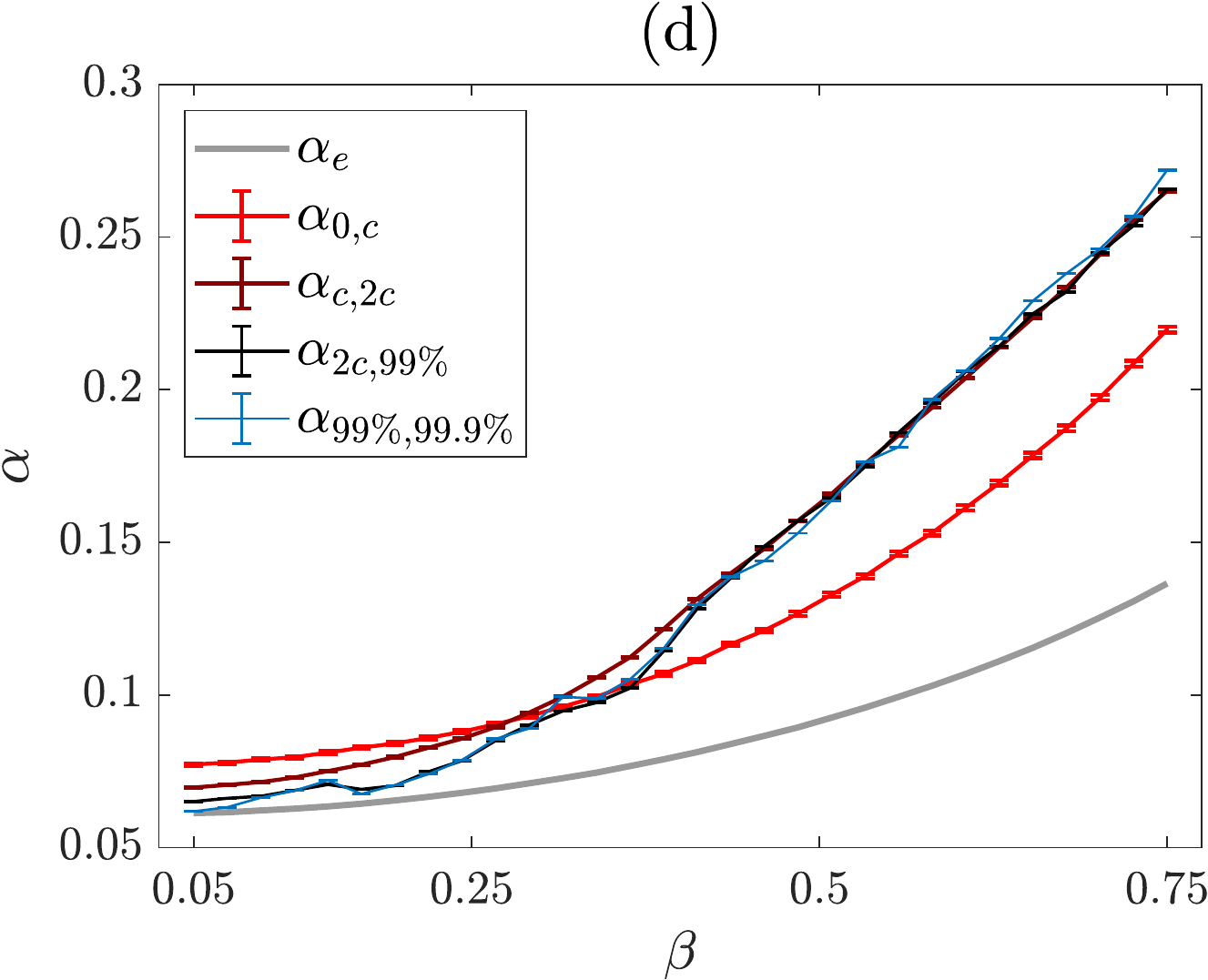}
	\caption{(a) Scheme of the Exit 1 for $\beta = 0.25$. (b) Hyperbolic particle decay for $\beta = 0.25$, $\beta = 0.5$ (solid lines) and $\beta = 0.75$ (dashed line). (c) Exponential fittings of the decay curves for only short times in the inertial relativistic approximation for $\beta \in [0.05, 0.75]$. (d) Similar fittings to (c) but for critical and  long times. We clarify that although the decay law $\alpha_e$ has been computed in terms of $\Delta E$, we show the results of fittings in terms of $\beta$ for simplicity (see Eq.~\eqref{eq:10}).}
	\label{fig:9}
\end{figure*}

In resemblance to the Newtonian case, the number of particles remaining inside the scattering region under hyperbolic dynamics at the time $t$ is $N(t) = N_0 e^{-\alpha_e t}$, where the $\alpha_e(\Delta E)$ is again the decay rate, but relativistic. It can be computed by setting that the flux of escaping particles through an exit of the well, for instance, the Exit 1, is \begin{equation} \iint dx d\theta \rho \textbf{v} \cdot \hat{\textbf{n}} = \int_{x_a}^{x_b} dx \int_0^\pi d\theta \rho(x,y,\theta) v(x,y) \sin \theta, \label{eq:13} \end{equation} where the quantity $\rho \textbf{v}$ means a current density vector and $\hat{\textbf{n}}$ is the normal vector to the Exit 1, which points outwards along the direction of the $y$-axis \cite{zhao2007}. We choose the Exit 1 for convenience, since the integral then remains defined along the segment delimited by $x \in [x_a,x_b]$ and $y = 1$, as shown in Fig.~\ref{fig:9}(a). Specifically, the points \begin{equation} x_{a,b} = \mp \sqrt{ \frac{2}{3} (\Delta E- c^2)} \label{eq:14} \end{equation} satisfy the condition $E = c^2 + V(x_{a,b},1)$, where the particle kinetic energy vanishes. Finally, bearing in mind the triangular symmetry of the well, the flux of escaping particles is the same through all exits, and therefore the decay rate of the system is three times the decay rate through the Exit 1, yielding the integral \begin{equation} \alpha_e(\Delta E) = \frac{3}{\pi c \Gamma(\Delta E)} \int_{-\sqrt{\frac{2}{3}(\Delta E - c^2)}}^{\sqrt{\frac{2}{3}(\Delta E - c^2)}} dx \sqrt{\left( \Delta E - \frac{3}{2}x^2 \right)^2-c^4}. \label{eq:15} \end{equation} It is possible to solve this integral by means of elliptic functions, obtaining the result \begin{equation} \alpha_e(\Delta E) = \frac{4 \sqrt{2} \Delta E \sqrt{\Delta E + c^2}}{\sqrt{3}\pi c \Gamma (\Delta E)} \left\{ {\rm E}\left( \frac{\Delta E - c^2}{\Delta E + c^2} \right) - \frac{c^2}{\Delta E} {\rm K}\left( \frac{\Delta E - c^2}{\Delta E + c^2} \right) \right\}, \label{eq:16} \end{equation} where E$(m)$ and K$(m)$ are the elliptic integrals of first and second kind, respectively, and $m$ means the squared elliptic modulus. In our case, $m=(\Delta E-c^2)/(\Delta E+c^2)$.

The results are similar to the Newtonian case already discussed. Firstly, as the value of $\beta$ is varied, the slope of the exponential decay curves becomes greater, as shown in Fig.~\ref{fig:9}(b). On the other hand, the shorter the fitting time of the decay curve is considered, the better the agreement between the numerical fittings and the ergodic decay rate $\alpha_e$, as displayed in Fig.~\ref{fig:9}(c). This is due to the fact that the chaotic saddle barely affects the trajectories escaping during the very first instants of the time evolution. Hence, as in the Newtonian case, the ergodic decay rate, $\alpha_e$, only takes into account the initial distribution of particles, in our case, randomly distributed inside the well. However, this initial distribution of particles is modified by the saddle since the time evolution begins. That is why only short-time fittings work. This last reasoning is also valid to interpret why the critical-time fittings do not strictly agree with the analytical decay rate, as seen in Fig.~\ref{fig:9}(d). Finally, when the relativistic system becomes very energetic, the ergodic decay rate underestimates all long-time fittings, $\alpha_{c,2c}$, $\alpha_{2c,99\%}$ and $\alpha_{99\%,99.9\%}$, similarly to the Newtonian case.

Moreover, we would like to mention that the formalism of ergodic decay laws discussed and developed along this work also fails when the regime is highly energetic (Newtonian and relativistic schemes included). In Fig.~\ref{fig:9}(b), for $\beta = 0.75$, it is noticeable that the first change of tendency in the decay curve happens before the indicated critical time (red triangle). Then, a problem arises, because the critical time is providing us misleading or unclear information regarding tendency changes. We argue that in very energetic cases the scattering region that we define, delimited by the potential saddle points, becomes far from the actual scattering region, delimited by the Lyapunov orbits, and does not reflect the actual phenomenology of escapes in the system. This produces the effect that the decay curve shows changes in the tendency before (and after as well) the critical time. In other words, when the system is highly energetic, the critical time becomes irrelevant to study the decay laws.

Time is absolute in nonrelativistic systems, but the measure of time in relativistic systems may depend on the choice of the reference frame. Escape times have been measured by an inertial clock at rest in the potential well in this work so far. However, it is possible to consider another noninertial reference frame comoving with a particle that describes a trajectory for a long time enough to measure the escape times of the other particles. The hypothetical clock attached to this particle will suffer the well-known time dilation phenomenon and measure the so-called proper time \cite{barton1999}, which in few words corresponds to the travel time of the twin who leaves the Earth in the twin paradox. Hence, given an infinitesimal time interval $dt$ as measured by an inertial clock, the particle clock will measure the time \begin{equation} d\tau = \frac{dt}{\gamma(t)}, \label{eq:17} \end{equation} where $\gamma(t)$ is the Lorentz factor at  time $t$. Both time intervals satisfy that $d\tau \le dt$, since the Lorentz factor is usually greater than unity along the particle trajectory. Importantly, note that it is necessary to know the complete Lorentz factor evolution of the particle clock to measure the proper time of an escaping trajectory, which is the result of integrating Eq.~\eqref{eq:17}.

Nonetheless, as the particle dynamics is bounded under the same energetic conditions, given a value of $\beta$, the Lorentz factor of all the trajectories is similar on average at any time $t$ before escaping. Then, there exists an approximate average value of the Lorentz factor and it can be reasonably estimated as the arithmetic mean between the maximum and the minimum values of the Lorentz factor inside the potential well, i.e., \begin{equation} \bar{\gamma}(\beta) = \frac{1 + \sqrt{1-\beta^2}}{2\sqrt{1-\beta^2}}, \label{eq:18} \end{equation} as introduced in \cite{fernandez2020}. Taking this into consideration, Eq.~\eqref{eq:17} can be approximated as $d\bar{\tau} \equiv dt/\bar{\gamma}$, where $\bar{\gamma}$ is a constant. Hence, escape times as measured by an inertial clock will be higher than escape times of the same trajectories as measured by the noninertial particle clock. This latter affects quantitatively the exponential decay laws. The decay rate in the inertial case, for instance, $\alpha_e$, will be smaller than the decay rate in the noninertial case, yielding \begin{equation} \bar{\alpha}^{\tau}_e \equiv \bar{\gamma} \alpha_e, \label{eq:19} \end{equation} where the approximation of bounded dynamics have been taken into account.

\begin{figure}[htp!]
	\centering
	
	\includegraphics[width=0.475\textwidth]{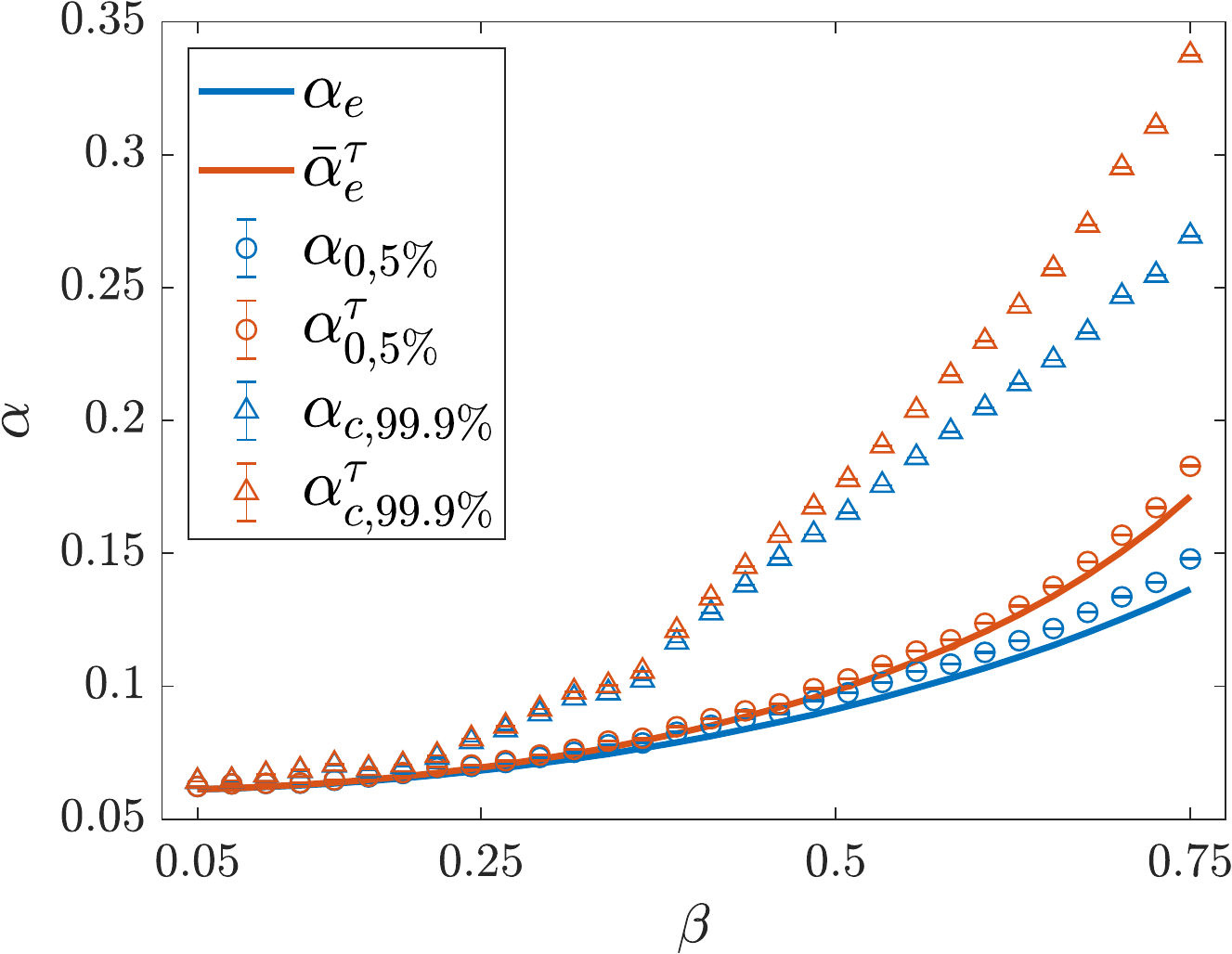}
	\caption{Exponential fittings for short times and long times to compare the decay rates as measured by an inertial clock at rest in the potential well in blue ($\alpha_e$, $\alpha_{0,5\%}$ and $\alpha_{c,99.9\%}$) and as observed from a noninertial particle clock evolving chaotically inside the scattering region in red ($\bar{\alpha}^{\tau}_e$, $\alpha^{\tau}_{0,5\%}$ and $\alpha^{\tau}_{c,99.9\%}$).}
	\label{fig:10}
\end{figure}

Observing particle escapes from a noninertial clock makes them occur faster due to the time dilation both at short and long times, according to the results shown in Fig.~\ref{fig:10}. Moreover, the fitted decay rate $\alpha^{\tau}_{0,5\%}$ for very short times agrees with the noninertial ergodic decay rate $\bar{\alpha}^\tau_{e}$. Thus, the approximation of assuming an average Lorentz factor inside the scattering region to characterize escapes works, at least for short times, which corresponds to the real range of applicability of ergodic decay laws, as proven in previous sections of this work.

\section{Conclusions} \label{sec:5}

The indiscriminate use of the concepts of hyperbolicity, chaos and ergodicity is sometimes found in the bibliography referring to dynamical systems with a few degrees of freedom, because chaos is usually responsible for the fulfillment of ergodicity in this kind of systems. In this manner, we have contextualized in the present work each of these terms in the open regime of the H\'{e}non-Heiles system and provided solid evidence that, far from being equivalent, chaos and the fact that the system is open can also be the precise reason why its motion is non-ergodic. In particular, we refer to the existence of a chaotic saddle, which provokes that particles explore all the regions of the energy surface unevenly before escaping.

Therefore, we have studied numerically an exponential decay law based on ergodic motions in the Newtonian H\'{e}non-Heiles system. We conclude that this ergodic decay law is able to quantify adequately escapes only for the very first instants of the time evolution when particles are launched randomly distributed. Thus, when the time evolution begins, particles remain no longer equally distributed throughout the scattering region, and in this way no longer conform to the microcanonical ensemble, on which the ergodic decay law is based. Indeed, particles evolve according to other more complex probability distributions given by the Liouville equation which dictates the evolution of the probability density according to the sets that govern the dynamics, i.e., KAM tori or the chaotic saddle in the nonhyperbolic and hyperbolic regimes, respectively.

In addition to the detailed study of the Newtonian case, we have extended the formalism of the ergodic decay laws to a relativistic version of the H\'{e}non-Heiles system for the first time. The result, obtained by means of measuring escapes from two inertial and noninertial reference frames, is that we have encountered the limitations present in the Newtonian case, for the same reasons. Nonetheless, the current work constitutes a step towards the solution in the characterization of decay laws in open systems. Bearing in mind the limitations of the microcanonical distribution, other distributions based on the sets that governs the dynamics have to be considered in the calculus of new decay laws. For instance, when KAM tori are present, the particle distribution must be affected by the phenomenon of stickiness. Therefore, in this case, there exist some regions near KAM tori where particles explore for longer times that others. On the other hand, when the chaotic saddle governs the dynamics, a fractalized particle distribution over the saddle can definitely describe all the escaping orbits in the hyperbolic regime. In general, we insist, the measure in the open regime evolves according to Liouville equation and converges to a fractal probability density defined on the chaotic saddle.

The critical time has to be considered as well, since it is part of the phenomenology caused by the invariant sets. Importantly, our numerical study has allowed us to interpret that the so-called critical time in open Hamiltonian systems means the maximum escape time that a trajectory is able to spend inside the scattering region without being affected at all by KAM tori or the chaotic saddle. Then, the existence of escaping orbits unaffected by them is reflected in the decay curves, which suffer a sudden change in their tendency when the critical time is reached. This fact opens some room to explore particle distributions that evolve in time.

To conclude, we mention some processes in nature with only a few relevant degrees of freedom where our numerical work can be useful. For example, open systems can be interpreted as scattering models of nuclear or chemical reactions, where each exit represents a final rearrangement of atoms and molecules \cite{zotos2015}. Interestingly, these kinds of reactions may occur chaotically and, in light of the present results, non-ergodically. Another relevant scattering phenomenon is the chaotic interaction that takes place among the resulting components of molecules above their dissociation energy. These processes constitute a sort of laboratory for studying the so-called quantum-classical correspondences (QCC) present in coupled nonlinear oscillators \cite{lin2013}, and perhaps our conclusions about ergodicity might be helpful to understand the formation of different bounded and unbounded states.

On the other hand, relaxation of nonequilibrium systems such as glass-forming melts or soft matter can be modeled by means of a modified H\'{e}non-Heiles system, whose saddle points are located at different values of the potential energy \cite{toledomarin2018}. More specifically, processes of relaxation consist of local reorientations of the masses that form the glass structure. In some cases, these processes are governed by a chaotic saddle, and hence new studies can be addressed to analyze how non-ergodic dynamics may also affect the structure and properties of glasses. Finally, the interest in diffusion phenomena in the so-called soft Lorentz gases has grown because these systems are able to simulate the electronic transport in graphene-like structures \cite{klages2019}. The unit cell of the triangular lattice of potentials in such systems is reminiscent to the morphology of the H\'{e}non-Heiles potential. Numerical works have evinced that soft Lorentz gases exhibit normal diffusion or superdiffusion due to a non-trivial interaction between trapped and ballistic periodic orbits that form the KAM tori. Superdiffusion is related to a dependence on initial conditions, and hence ergodicity is broken in such a case. Thus, the present work can provide new approaches to elucidate this phenomenology.

\section*{ACKNOWLEDGMENTS}

This work has been supported by the Spanish State Research Agency (AEI) and the European Regional Development Fund (ERDF, EU) under Projects No.~FIS2016-76883-P and No.~PID2019-105554GB-I00.
	
\appendix*
\section{Comments on computing the fraction of KAM tori in a surface of section.} \label{sec:6}
	
As indicated, an ensemble of particles are launched from the exits of the scattering region with random initial shooting angles into its interior. Such initial conditions, blue, red and green colored in Fig.~\ref{fig:2}(a), are \textit{all} the possible final conditions of trajectories just before escaping forever, but with their velocity vector rotated by $\pi$ radians. For clarity, the limit case of \textit{all} possible final conditions is reached when the number of initial conditions is infinite. Thus,  due to the time symmetry of conservative systems, we simulate \textit{all} escaping trajectories but backwards in time.

Another basic method to compute the fraction of KAM tori consists of counting how many trajectories do not escape of the total number of trajectories initially launched \cite{nieto2020}. Such a method is usually computationally expensive as it has to evolve trajectories belonging to KAM tori, which do not escape by definition, in addition to trajectories with long transients due to KAM stickiness. The latter trajectories, if the final computation time is less than their escape times, can be interpreted as belonging to KAM tori, which would yield an erroneous result of the fraction of KAM tori. Nonetheless, the alternative method proposed here depends on the number of particles launched: the greater the number of escaping particles, the greater the probability of mapping the whole area of the surface of section occupied by the chaotic regions. Specifically, the regions of the surface of section that are most complicated to map are those away from the fractal boundaries of exit basins and tori KAM. These regions, which grow as the energy of the system increases, are formed by a multitude of trajectories that we must launch, which escape without exhibiting long transients and map only a few points on the surface of section.

For completeness, we add two more method's aspects. On the one hand, note that it is totally equivalent launching an infinite ensemble of particles by the proposed method that simulating a single particle trajectory that elastically bounces off the exits forever \cite{zheng1995}, since in both cases one would obtain the same fraction of KAM tori. On the other hand, this computationally affordable method can be applied to any system (including area-preserving maps and continuous-time systems) that exhibits exits well-defined where particles can be started from.

\end{document}